\documentclass[11pt]{iopart}
\usepackage{graphicx}
\usepackage{url}
\usepackage{comment}

\usepackage{xcolor}

\begin{document}

\title[S/XB from W$^{5+}$ lines]{Evaluation of tungsten influx rate using 
	line emissions from W$^{5+}$ ions in EAST Tokamak}
	
\author{Fengling Zhang$^{1,2}$, Dar\'{\i}o Mitnik$^{*3,1}$,
Ling Zhang$^{\dag 1}$, Runjia Bao$^{1,2}$,
Wenmin Zhang$^{1,2}$, Shigeru Morita$^{4,1}$,
Yunxin Cheng$^{1}$, Ailan Hu$^{1}$,
Chengxi Zhou$^{1}$, Jihui Chen$^{1}$, 
Xiaobin Ding$^5$, Yinxian Jie$^{1}$, and Haiqing Liu$^{1}$}

\address{$^1$ Institute of Plasma Physics, Hefei Institutes of Physical Science, Chinese Academy of Sciences, Hefei 230031, China}
\address{$^2$ University of Science and Technology of China, Hefei 230026, China}
\address{$^3$ Instituto de Astronomía y Física del Espacio (CONICET-Universidad de Buenos Aires), Buenos Aires 1428, Argentina}
\address{$^4$ National Institute for Fusion Science, Toki 509-5292, Gifu, Japan}
\address{$^5$ Key laboratory of Atomic and Molecular Physics
$\&$ Functional Materials of Gansu Province, College of Physics and Electronic Engineering, Northwest Normal University, Lanzhou, 730070, China}

\ead{$^{*}$dmitnik@df.uba.ar, $^{\dag}$zhangling@ipp.ac.cn}

\vspace{10pt}
\begin{indented}
\item[]September 2024 
\end{indented}

\begin{abstract}
The S/XB ratios (ionization events per emitted photon) allow one to relate 
spectroscopic emissivity measurements to the impurity influx 
from a localized source. 
In this work, we determine the tungsten influx by examining two 
dominant EUV (Extreme Ultraviolet) line emissions at 382.13 \AA~and 394.07 \AA, 
corresponding to the $4f^{14}5f \rightarrow 4f^{14}5d$ radiative transitions 
of the W$^{5+}$ ion. 
The ground configuration of W$^{5+}$ consists of the ground 
level and a metastable level, with the latter having a higher population than the ground state.
Therefore, a simple approach assuming that the transitions are independent, i.e., 
only populated by a unique level source, requires correction. 

To address this, we have developed a fully collisional–radiative modeling in which 430 levels contribute to the ionization. We have utilized three advanced computational codes -- {\sc hullac} (Hebrew University - Lawrence Livermore Atomic Code), {\sc as} (AutoStructure), and {\sc fac} (Flexible Atomic Code) -- for the atomic structure calculations. These codes provide the necessary information such as wavelengths, collisional and radiative transition rate coefficients. The {\sc fac} code was also used to calculate the direct electron-impact ionization under the distorted-wave approximation. 
We also included contributions to total ionization from excitation-autoionization processes up to $n=15$ manifolds from the distorted-wave calculations. 

Subsequently, we used these results to ascertain the tungsten impurity influx in a dedicated discharge of the EAST tokamak, which operates with full tungsten divertors. In our findings, we observed that for the density range relevant to the edge region of a tokamak plasma, the S/XB ratios are almost independent of electron density but exhibit significant variation with electron temperature. 	
\end{abstract}
\vspace{2pc}

\maketitle


\section{INTRODUCTION}

Various intrinsic impurities are inevitably generated in tokamak 
discharges due to the interaction between plasma and wall. 
Impurities ionized in the core release a large number of electrons, severely diluting the concentration of the primary plasma and affecting the reaction power density of fusion,  
decreasing the overall performance of the plasma. 
Therefore, monitoring and controlling the entry of impurities into the core are an important issue for long-pulse discharge operations. 

It is possible to relate spectroscopic measurements of emissivities 
from an impurity ion to the impurity influx into the plasma core. 
Behringer {\it et al.} \cite{Behringer:89} proposed the use of the 
``ionization per emitted photon" S/XB ratio, also known as the 
``inverse photon efficiency", denoting the number of ionization 
events per observed photons (incoming ions over emitted photons).
The S/XB ratios have been used to evaluate the influx in 
tokamak devices of common light impurities such as carbon, nitrogen, oxygen, and neon, as well as metal impurities such as chromium, iron, nickel, molybdenum, and others \cite{Behringer:89,Field:96,Badnell:96,Griffin:97,Lipschultz:01}.

Tungsten has been selected as the divertor and first wall material for ITER due to its high melting point, high thermal conductivity, low sputtering rate, low tritium retention rate, and low neutron activation rate \cite{Coenen:15}. 
However, highly ionized tungsten ions in the core and edge emit intense radiation at various wavelength ranges, which can cause significant radiation power loss and severely reduce plasma confinement performance. Therefore, evaluating the amount of tungsten entering through the edge is crucial for functioning high confinement long pulse plasma. 
Many extensive studies have been dedicated to the evaluation of the tungsten influx using the 
4009 \AA~spectroscopic line of W I \cite{Thoma:97,Geier:02,Nishijima:11}. 
Other lines from W I have been used by Beigman {\it et al.} \cite{Beigman:07}, and W II was used by Pospieszczyk {\it et al.} \cite{Pospieszczyk:10}.
Ballance {\it et al.} \cite{Ballance:13} conducted a detailed study on 
the S/XB ratios for the 1099.05 \AA, 1119.7 \AA, 1172.47 \AA, and 1186.17 \AA~lines of W$^{3+}$, 
assuming these lines as independent.
Dong {\it et al.} \cite{Dong:19} evaluated the influx rate of W$^{6+}$ using the S/XB 
coefficients calculated with the ADAS code \cite{ADAS}.
 
Recently, the spectrum of W$^{5+}$ was observed during EAST discharge, which provides us with the necessary experimental spectrum for further studying the S/XB of low-ionized tungsten ions.
As pointed out by Behringer {\it et al.} \cite{Behringer:89}, data for any particular ionization stage of the atomic species of interest can be employed as long as no higher ionization stages of that atom emerge from the localized source. 
Thus, such spectroscopic measurements for specific radiative transitions in W$^{5+}$  would allow for the determination of the influx of tungsten.

The most straightforward approach for the S/XB determination is to assume that 
only one ground (or metastable) level populates the upper level of the observed 
spectral line. 
Even if the effects of collisions to and cascades from higher lying levels are considered, this simple approximation only works for some cases, as shown in the following sections. In general, we must take allowance that the source of the level population can also arise from other metastables and not only from a unique level, the so-called {\it metastable cross-coupling approach} \cite{Badnell:96}.

The paper is organized as follows. 
In Section~\ref{sec:theory}, we develop the main formulas needed for the 
evaluation of the S/XB coefficients. 
First, we show in \ref{subsec:CR} details about how to employ the 
rate coefficients for excitations and radiative decays into 
a quasistatic generalized collisional–radiative level population model. 
With this, we obtain the effective contribution 
to the population of the excited states involved in the radiative 
transitions to study. 
In \ref{subsec:SXB}, we show how to use these coefficients 
to determine S/XB ratios as a function of electron
density and temperature.
We realized that many works deal with cases in which only the ground level is populated or cases in which the emissivity in the metastable line could be considered independent (namely, the upper levels are only populated by direct excitation from a unique source).   
However, the discussion for cross-coupled metastables, in which the populations of the upper levels arise from different sources, is scarce and worth developing in detail.
In Subsection \ref{subsec:calcmethods}, we briefly describe the three computational codes used in our calculations for the atomic structure.
In Section \ref{sec:results}, we show the results, first for the atomic structure calculations in  \ref{subsec:atomic}.
Then, we present the results obtained for a model consisting of only 
four levels, which accounts for most of the relevant physical processes 
involved, at least within the density and temperature range of interest 
for Tokamak plasmas. 
Following that, we solve a full collisional-radiative model with 430 levels, 
and present and discuss the final S/XB coefficients.
We conclude with some final remarks in Section~\ref{sec:conclusions}.

\section{Experimental setup}
\label{sec:experimental}

EAST is a fully superconducting tokamak device equipped with ITER-like active water-cooled W/Cu monoblocks tungsten divertors, capable of high-power long pulse operation. The main parameters are: major radius $ R \leq 1.9$ m, 
minor radius $a \leq 0.45$ m, plasma current $I_p \leq 1$ MA, and 
toroidal magnetic field $B_T \leq 3.5$ T \cite{Wan:17,Wan:19,Wang:21}. 
Presently, various auxiliary heating and current driving have been installed, including two lower hybrid current drive (LHCD) systems, an electron cyclotron heating (ECH) system, an ion cyclotron resonant frequency (ICRF) system, and balanced neutral beam injection (NBI) systems \cite{Wan:22}.
A set of flat-field extreme ultraviolet (EUV) spectrometers, working in the 20-500 \AA~wavelength range with fast time response (5 ms/frame), named EUV\_long \cite{ZhangLing:15}, was developed to measure line emissions and to monitor the impurities present in the EAST plasmas.
A laminar-type varied-line-spacing concave holographic grating 1200 grooves/mm is fixed at grazing incidence $87^\circ$ with a narrow entrance slit width of 30 $\mu$m for increasing the spectral resolution. 
A back-illuminated charge-coupled device (CCD) with a total size of 26.6 × 6.6 mm$^2$ and pixel numbers of $1024 \times 255$ 
($26 \times 26$ $\mu$m$^2$/pixel) is used for recording the spectrum. 
The wavelength calibration is performed by cubic polynomial fitting with many well-known spectral lines covering the observable range \cite{Lei:21}.
An absolute calibration, in which the raw measured spectral counts 
are converted into absolute spectral intensities, 
is further achieved by comparing the observed and calculated intensities of 
EUV and visible bremsstrahlung continua, a procedure explained in 
previous works \cite{ZhangLing:15,Lei:21,Dong:11,Dong:12}. 

\begin{center}
\begin{figure}[h!]
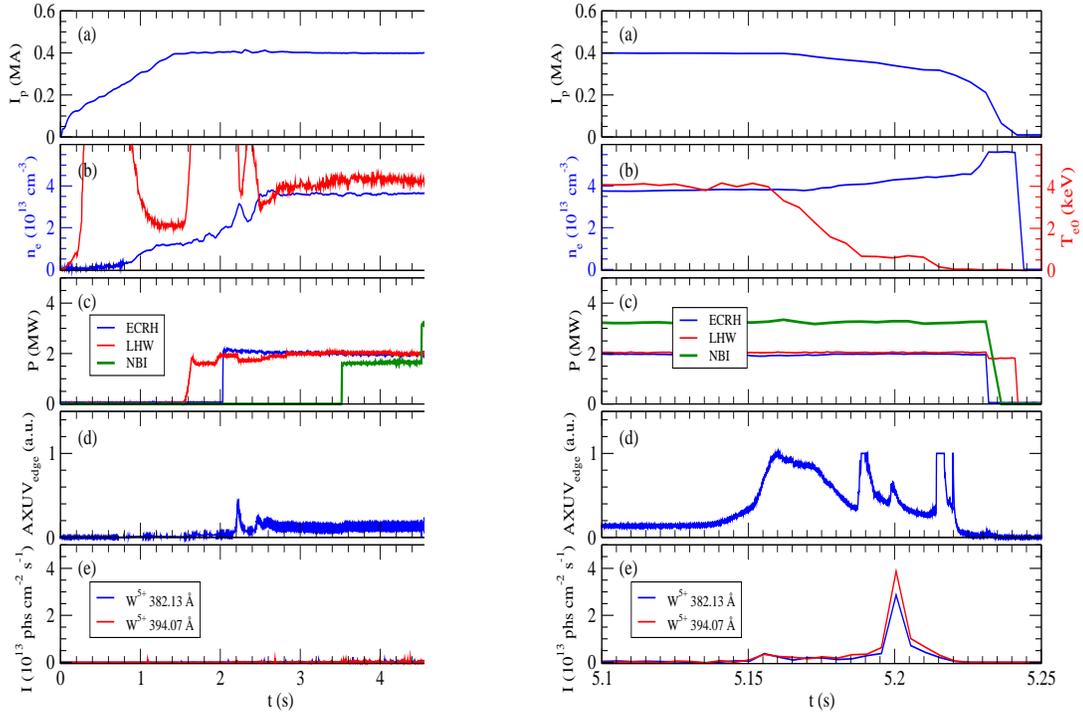

\hfill 
\includegraphics[width=0.45\textwidth,height=0.4\textheight]{mdsfull.eps}
\includegraphics[width=0.45\textwidth,height=0.4\textheight]{mds_new.eps}  
\caption{Evolution of specific parameters for the EAST discharge \#100300. 
(a) Plasma current $I_p$. 
(b) Central electron temperature, $T_{e0}$, and line-averaged electron density, $n_e$. 
(c) Heating power of ECRH, LHW, and NBI. 
(d) Boundary radiation power loss. 
(e)  W$^{5+}$ emission line intensities at 382.13 \AA~(blue), and 394.07 \AA~(red).
Right Part: Zoom at $t=5.10-5.25$ s.}
\label{fig:mds}
\end{figure}
\end{center}

Figure~\ref{fig:mds} (left part) displays the operating parameters of EAST plasma discharge \#100300 as a function of time. The upper part (a) displays the plasma current $I_p$. Part (b) presents the electron temperature of the core plasma $T_{e0}$ provided by the electron cyclotron emission (ECE) system \cite{Han:18} and the line-averaged electron density $n_e$ measured by the hydrogen cyanide (HCN) interferometer system \cite{Xu:08}. 
Part (c) depicts the heating power of ECRH, LHW, and NBI. 
Part (d) shows the total radiation power loss measured by the AXUV (absolute extreme ultraviolet) photodiode array \cite{Duan:11}.
Finally, part (e) presents the emission intensity of the two dominant W$^{5+}$ EUV spectral lines at 382.13 \AA~and 394.07 \AA, which are used here to determine the total flux of tungsten. These two lines are not frequently observed in general discharges. As the figure indicates, they do not appear during the initial and flat-top stages of the discharge, likely due to a significant reduction in throughput at longer wavelengths in the current grazing-incidence EUV spectrometer.
However, at $t \approx 5.2$ s, high-intensity W$^{5+}$ lines are detected. This is attributed to strong plasma-wall interactions (PWI) that ultimately lead to plasma termination. 
Therefore, we focus on the plasma behaviour at this final stage of the discharge, during the time interval $t=5.10-5.25$, as depicted in the right part of Figure~\ref{fig:mds}. 
As shown in part (d), the radiation power loss begins to increase at time $t=5.14$ s due to core tungsten accumulation. This accumulation causes the plasma to cool down dramatically, as illustrated in part (b). 
When the electron temperature is relatively low, for instance, when $T_{e0} \leq 1.0$ keV,
the emission volume of the W$^{5+}$ lines increases, and meanwhile, the tungsten source increases mainly due to this strong PWI, then high-intensity spectral lines of W$^{5+}$ at 382.13 \AA~and 394.07 \AA~could be observed, as shown in part (e).

\begin{center}
\begin{figure} 
\centering\includegraphics[width=1.0\textwidth]{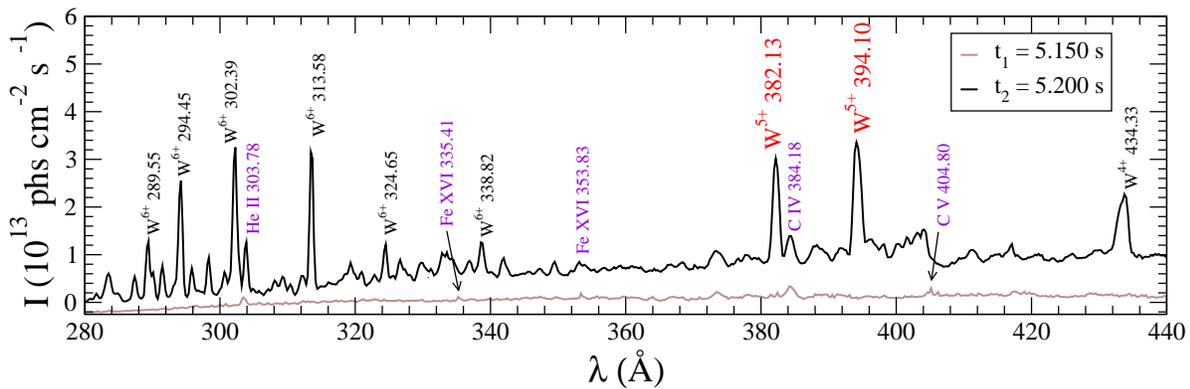}
\caption{EAST spectra obtained at discharge \#100300 with the 
b\_long spectrometer. 
Brown curve: Observed at $t=5.150$, before the W burst.
Black curve: Observed at $t=5.200$, at the maximum of W intensity.}
\label{fig:rawspectra}
\end{figure}
\end{center}

Figure~\ref{fig:rawspectra} displays the 280-440 \AA~spectral range 
for the EAST discharge \#100300 at two different times. 
The brown curve was obtained at time $t=5.150$ s, when   
no trace of W was found in the plasma, and  
the black curve was obtained at time $t=5.200$ s when 
the W intensity reached its maximum. 
Before the burst, only some lines emitted from HeII, Fe XVI, CIV, and 
CV are observable in this spectral range. 
The lines observed at the W peak intensity have been identified \cite{Zhangfl:24,Zhangwm:24}
as belonging to different low-charged W ions, as indicated in the figure. 
The strong lines in the 280-320 \AA~range are associated with the 
$5p^6 \rightarrow 5p^55d$ transitions of W$^{6+}$, and from this ion,  
the $4f^{13}5s^25p^65d \rightarrow 4f^{13}5s^25p^65f$ lines are also 
observed. 
The line at 434.33\AA, corresponds to the $5d^2 \rightarrow 5d7p$ 
transition of W$^{4+}$. 
In the region around 400 \AA, two prominent lines are highlighted. 
We have identified these lines as corresponding to $5f \rightarrow 5d$  
transitions of W$^{5+}$; the line at 382.13 \AA~is emitted by the 
$(4f^{14}5f)_\frac{5}{2} \rightarrow (4f^{14}5d)_\frac{3}{2}$
transition, and the line at 394.07 \AA~arises from the 
$(4f^{14}5f)_\frac{7}{2} \rightarrow (4f^{14}5d)_\frac{5}{2}$
transition.
As we shall see later, these lines will provide the spectroscopic 
information necessary to determine the amount of tungsten flux entering 
the plasma.

\bigskip

\clearpage
\pagebreak
\section{Theoretical methods}
\label{sec:theory}

\subsection{Collisional Radiative Model}
\label{subsec:CR}

To determine the impurity influx for a particular element using
the photon emission from a particular ion, one needs to know the
population of the excited levels of this ion.
To calculate the population $n_j$ of the excited levels $j$ of a 
particular impurity ion, we solve for a given electronic density and 
temperature, the collisional-radiative equations:
\begin{eqnarray}
\frac{d n_j}{dt} &=& 
\sum_{k>j} n_k \left(A_{kj} + n_e \, Q_{kj} \right) + 
\sum_{i<j} n_i \, n_e \, Q_{ij}  - 
\nonumber \\
& & - 
n_j \left(  \sum_{i<j}    \left(A_{ji} + n_e \, Q_{ji} \right) +  
\sum_{k>j} n_e \, Q_{jk}  \right) \,\, = 
\nonumber \\
&=&
\sum_{m} C_{jm} \, n_m  \, ,
\label{eq:crm} 
\end{eqnarray}
where $n_e$ is the electronic density, 
$A_{ji}$ are the radiative rate coefficient for transitions 
from level $j$ to level $i$.  
$Q_{ji}$ are the electron-impact excitation rate coefficients if $i>j$, 
and the deexcitation rates if $i<j$.
In matrix form, Eq.~(\ref{eq:crm}) is 
\begin{eqnarray}
\frac{d \vec{n}}{dt} = {\mathbf C} \cdot  \vec{n}
\label{eq:crmatrix}
\end{eqnarray}
where the elements of the matrix ${\mathbf C}$ are
\begin{eqnarray}
C_{jm} \equiv 
\left\{
\begin{array}{cc}
A_{mj} + n_e \, Q_{mj} &  ~~~~\mathrm{for~} m>j \\
         n_e \, Q_{mj} &  ~~~~\mathrm{for~} m<j \\
-\left(  \sum_{i<j}    \left(A_{ji} + n_e \, Q_{ji} \right) +  
\sum_{k>j} n_e \, Q_{jk}  \right) &  ~~~~\mathrm{for~} m=j
\end{array}
\right.
\label{eq:Ccoeff}
\end{eqnarray}
We partition the complete set of levels into metastables 
(denoted by greek letters $\sigma,\rho,\ldots,\mu$), and 
excited levels (denoted here with roman letters 
$i,j,\ldots,n$) 
\begin{equation}
\vec{n} = 
\left(
\begin{array}{c}
\vec{n}_\tau \\
\vec{n}_s
\end{array}
\right)
\end{equation}
and assume a quasi--static approximation in which the 
population of the excited levels are relaxed relative to the ground 
and metastable levels:
\begin{equation}
\frac{d \vec{n}_\tau}{dt} \ne 0 \hfill \mathrm{and} 
\hfill \frac{d \vec{n}_s}{dt}=0  \, . \hfill
\end{equation}
Under this approximation, Eq.~(\ref{eq:crmatrix}) becomes

\begin{eqnarray}
\left(
\begin{array}{c}
~ \\
\frac{d \vec{n}_\tau}{dt} \\
~ \\
\hline
~ \\
~ \\
 \frac{d \vec{n}_s}{dt}=0 \\
~ \\
~ \\
\end{array}
\right)
=
\left(
\begin{array}{ll|ccccc}
    &   &          &  &   &  & \\ 
    &   &          &  &   &  & \\ 
    &   &          &  &   &  &  \\
    \hline
    &   &          &  &   &  &  \\ 
    &   &          &  &   &  &  \\ 
    &{\mathbf C_{\tau}}~  &          &  & {\mathbf C_{s}}  &  &   \\ 
    &   &          &  &   &  &  \\ 
    &   &          &  &   &  &  
\end{array}
\right)
\left(
\begin{array}{c}
~ \\
\vec{n}_\tau  \\
~ \\
\hline
~ \\
~ \\
\vec{n}_s  \\
~ \\
~ \\
\end{array}
\right)
\end{eqnarray}

To solve this problem, we take only the excited-levels part of 
the matrix, having the following equation
\begin{eqnarray}
0 &=& {\mathbf C_\tau} \cdot  \vec{n}_\tau + {\mathbf C_s} \cdot  \vec{n}_s 
\end{eqnarray}
The solution for the excited levels is
\begin{eqnarray}
\vec{n}_s & = & 
[{\mathbf C_s}]^{-1} \cdot  [-{\mathbf C_\tau}] \cdot  \vec{n}_\tau 
\equiv
n_e \, {\mathbf {\cal F}} \cdot \vec{n}_\tau \, .
\end{eqnarray}
The matrix ${\mathbf {\cal F}}$ denotes the effective contribution to the population of the excited levels via excitations from the metastables. 
This is the usual notation used in the field, in which ${\cal F}_{i \rho}$ represents the effective contribution to the population of $i$ due to excitations from the metastable $\rho$.

\pagebreak
\subsection{S/XB --  Ionization Events per Photon}
\label{subsec:SXB}

The neutral impurity particles entering the Tokamak plasma from 
the wall are ionized in a narrow region around the surface. 
Neglecting recombination processes, the inward impurity influx 
density is equal to the ionization rate per unit surface area, 
integrated over the extension of the region towards the 
center of the plasma. 
Behringer  {\it et al.} have shown  \cite{Behringer:89} 
that for the calculation of the inward flux, it is enough to 
choose an ionization degree $Z$ of the impurity ions sufficiently small 
to ensure that no ions emerge from the sputtering surface in higher 
ionization stages.
In that case, the inward flux along the line-of-sight $\xi$ up to stage $Z$
becomes the overall impurity flux, which is related to the 
integral over the abundances of the metastables of stage $Z$ alone:
\begin{eqnarray}
\Gamma = \Gamma_\sigma &=& \int_0^{\infty} 
n_e \,\, \sum_\sigma S_\sigma(\xi) \,\, n_\sigma(\xi) \, d\xi \,,
\label{eq:flux}
\end{eqnarray}
where $S_{\sigma}$ is the ionization rate coefficient from 
metastable level $\sigma$ of the impurity ion of charge $Z$.
Since a ionization event is followed by the 
emission of spectral lines, it is possible to relate the number of 
emitted photons with the flux.
Considering the radiative transition from a level $j$ to a lower level $i$  
and supposing that the upper-level $j$ is populated only from levels 
excited from a unique metastable $\sigma$, 
the strength of a spectral line produced by the radiative transition 
$j \rightarrow i$ is proportional to its emissivity
\begin{eqnarray}
\epsilon_{\sigma,i\rightarrow j} = A_{ij} \, n_i
\end{eqnarray}
and the corresponding line-of-sight emissivity is defined by
\begin{eqnarray}
I_{\sigma,i\rightarrow j} = \int_0^{\infty} \, 
\epsilon_{\sigma,i\rightarrow j}(\xi) \,\, d\xi \,. 
\end{eqnarray}
Substituting in the flux equation (\ref{eq:flux}) and taking into 
account that for only one metastable 
\begin{equation}
{\cal F}_{i\sigma}  = \frac{1}{n_e} \frac{n_i}{n_\sigma}
\label{eq:F}
\end{equation}
the flux becomes \cite{Behringer:89}:
\begin{eqnarray}
\Gamma_\sigma &=& \int_0^{\infty} 
n_e \,\, S_\sigma(\xi) \,\, n_\sigma(\xi) \, d\xi \approx 
\frac{S_\sigma}{A_{ij} \,\, {\cal F}_{i\sigma} } 
\,\, \int_0^{\infty} \, A_{ij} \,\, n_i(\xi) \, d\xi 
= \nonumber \\
&=& \frac{S_\sigma}{A_{ij} \,\, {\cal F}_{i\sigma} } 
\,\, I_{\sigma, i \rightarrow j} =  
{\cal SXB}_{\sigma, i \rightarrow j} \,\, I_{\sigma, i \rightarrow j} 
\, .
\label{eq:GammaF}
\end{eqnarray}
This equation allows us to identify the Ionization Events per Photon -- 
coefficient S/XB (for only one metastable $\sigma$) as:
\begin{equation}
{\cal SXB}_{\sigma, i \rightarrow j} =  
    \frac{   S_{\sigma}   }{ A_{ij} \, {\cal F}_{i\sigma} } \, .
\label{eq:SXBgs}
\end{equation}

Equation (\ref{eq:SXBgs}) implicitly captures the general dependence of the S/XB coefficients on plasma parameters. As the name suggests (and will be shown below), the S/XBs are proportional to the ionization rates and inversely proportional to the excitation rates. This roughly determines their dependence on electronic temperature: while both rates increase with temperature in the relevant ranges, the effect of ionization on temperature is more pronounced. Consequently, the coefficients tend to exhibit a monotonic increase with temperature. Concerning their dependence on electron density, the S/XB coefficients appear to be independent of density at low values. At intermediate density values, the relationship becomes cumbersome, as the excitations occur through an effective rate $\mathcal{F}$ described in expression (\ref{eq:F}), which requires solving a complex collisional-radiative problem that accounts for all cascade processes.
\subsection{Derivation of S/XB formula for many metastables}

Transitions from metastables can compete with the direct 
excitation from the ground state as the population processes 
for the upper transition levels. That happens even for low 
electronic densities and also even for small metastable populations.
Therefore, we need to reconsider Eq.~(\ref{eq:GammaF}) for {\it 
metastable cross-coupling}. 
First of all, if we have $\tau$ metastables, we need to measure 
the emissivity of $(n=\tau)$ different lines 
\begin{eqnarray}
&& I_1 = I_i \equiv I_{\sigma\rho\cdots\tau,i \rightarrow j} =
\int_0^{\infty} A_{ij} \,\, n_i(\xi) \, d\xi \nonumber \\
&& I_2 = I_l \equiv I_{ \sigma\rho\cdots\tau, l \rightarrow m} = 
\int_0^{\infty} A_{lm} \,\, n_l(\xi) \, d\xi \nonumber \\ 
&& ~~~~~\cdots \nonumber \\
&& I_n = I_s \equiv I_{ \sigma\rho\cdots\tau, s \rightarrow t} = 
\int_0^{\infty} A_{st} \,\, n_s(\xi) \, d\xi \nonumber   
\end{eqnarray}
in which the subscripts $\sigma\rho\cdots\tau$ indicate that,  
in principle, all the upper levels $i,l,\cdots,s$ can be populated by 
excitations from all the metastables, namely,
\begin{eqnarray}
n_i &=& n_e \,\, n_\sigma \,\, {\cal F}_{i\sigma} + 
	n_e \,\, n_\rho \,\, {\cal F}_{i\rho} + \cdots +
	n_e \,\, n_\tau \,\, {\cal F}_{i\tau}
\nonumber \\
n_l &=& n_e \,\, n_\sigma \,\, {\cal F}_{l\sigma} + 
	n_e \,\, n_\rho \,\, {\cal F}_{l\rho} + \cdots +
	n_e \,\, n_\tau \,\, {\cal F}_{l\tau}
\nonumber \\
&& ~~~~~\cdots 
\nonumber \\
n_s &=& n_e \,\, n_\sigma \,\, {\cal F}_{s\sigma} + 
	n_e \,\, n_\rho \,\, {\cal F}_{s\rho} + \cdots +
	n_e \,\, n_\tau \,\, {\cal F}_{s\tau}
\label{eq:niF}
\end{eqnarray}
which, in matrix form is written as:
\begin{eqnarray}
\frac{1}{n_e} \left( \vec{n}_k \right) = 
\left(   \mathbf{F}   \right) \, \, 
\left( \vec{n}_\mu \right) \, ,
\end{eqnarray}
where the reduced $n$-dimensional column vectors and the corresponding 
$n \times n$ operators are  
\hfill
\begin{eqnarray}
\left( \vec{n}_k \right) \equiv 
\left(
\begin{array}{c}
n_i \\ 
n_l \\
\cdots \\
n_s
\end{array}  
\right) ; \, 
\hfill  \,  
\left( \vec{n}_\mu \right) \equiv 
\left(
\begin{array}{c}
n_\sigma \\ 
n_\rho \\
\cdots \\
n_\tau
\end{array}  
\right) ; \,
\hfill \,  
\left(   \mathbf{F}   \right) \equiv 
\left(
\begin{array}{cccc}
{\cal F}_{i\sigma} & {\cal F}_{i\rho} & \cdots & {\cal F}_{i\tau} \\ 
{\cal F}_{l\sigma} & {\cal F}_{l\rho} & \cdots & {\cal F}_{l\tau} \\ 
\cdots             &\cdots            &\cdots  &\cdots             \\
{\cal F}_{s\sigma} & {\cal F}_{s\rho} & \cdots & {\cal F}_{s\tau} 
\end{array}  
\right)  
\hfill
\, .
\end{eqnarray}
\hfill

The population numbers of the metastable levels $n_\sigma$ and 
$n_\rho$ are: 
\begin{eqnarray}
\left( \vec{n}_\mu \right) = 
\frac{1}{n_e}  \, \left(   \mathbf{F}   \right)^{-1} \, \, 
\left( \vec{n}_k \right) \, 
= 
\frac{1}{n_e}  \, \left(\mathbf{R}\right)  \, \, 
\left( \vec{n}_k \right) \, ,
\label{eq:nmu}
\end{eqnarray}
in which the inverse of the reduced matrix $(\mathbf{F})$ is denoted as 
\begin{equation*}
(\mathbf{R}) \equiv (\mathbf{F})^{-1}
\end{equation*}

The total flux is 
\begin{eqnarray}
\Gamma &=& \Gamma_\sigma + \Gamma_\rho + \cdots + \Gamma_\tau =
\nonumber \\
&\approx&
n_e \, S_\sigma \int_0^\infty n_\sigma  + n_e \, S_\rho \int_0^\infty n_\rho 
+ \cdots + n_e \, S_\tau \int_0^\infty n_\tau = 
\nonumber \\
\nonumber \\
&=&
S_\sigma \int_0^\infty 
\left( {\cal R}_{\sigma i} n_i + {\cal R}_{\sigma l} n_l + \cdots + 
{\cal R}_{\sigma s} n_s \right) + 
S_\rho \int_0^\infty 
\left( {\cal R}_{\rho i} n_i + {\cal R}_{\rho l} n_l + \cdots + 
{\cal R}_{\rho s} n_s \right) + 
\nonumber \\
& & + \cdots + 
S_\tau \int_0^\infty 
\left( {\cal R}_{\tau i} n_i + {\cal R}_{\tau l} n_l + \cdots + 
{\cal R}_{\tau n} n_n \right) = 
\nonumber \\
&\approx&
\frac{1}{A_{ij}} \left(
S_\sigma \, {\cal R}_{\sigma i} + S_\rho \, {\cal R}_{\rho i} + \cdots +
S_\tau\, {\cal R}_{\tau i} \right) \times \int_0^\infty A_{ij} \, n_i +
\frac{1}{A_{lm}} \sum_{\mu=\sigma}^{\tau} 
S_\mu \, {\cal R}_{\mu l} \times I_l + \cdots +
\nonumber \\
& & +
\frac{1}{A_{n}} \sum_{\mu=\sigma}^{\tau} S_\mu \, 
{\cal R}_{\mu n } \times I_n = 
\sum_{k=1}^{n} {\cal SXB}_k \times I_k 
\label{eq:totalG}
\end{eqnarray}
where $A_n \equiv A_{st}$ and 
\begin{eqnarray}
{\cal SXB}_k \equiv 
\sum_\mu {\cal SXB}_{\mu k} = 
\frac{1}{A_{k}} \sum_{\mu} S_\mu \, {\cal R}_{\mu k } = 
 \frac{1}{A_{k}} \sum_{\mu} S_\mu \, ( \mathbf{F} )^{-1}_{\mu k } \,  .
\label{eq:sxbn}
\end{eqnarray}

\subsection{Calculational Methods}
\label{subsec:calcmethods}

We carried out an evaluation of the atomic structure, as well as the collisional and radiative rates required to solve the collisional-radiative problem, using three separate independent calculations. Specifically, we utilized the atomic structure codes {\sc hullac} (Hebrew University - Lawrence Livermore Atomic Code) \cite{Klapisch:77,BarShalom:01}, the {\sc as} (AutoStructure) code \cite{Badnell:86,Badnell:97,Badnell:11}, and the {\sc fac} (Flexible Atomic Code) \cite{Gu:08}. Since these three codes are well-established in both astrophysical and fusion research, we will only provide a brief overview of their key features.

The {\sc hullac} code is a fully relativistic atomic structure package that computes transition energies and corresponding rate coefficients, among other quantities. It employs a relativistic potential method to solve the Dirac Hamiltonian and uses full multiconfiguration wavefunctions to calculate radiative transition rates. Configuration mixing, and hence correlation effects, are included in the calculations. The Breit interaction and quantum electrodynamics corrections (vacuum polarization and self-energy) are treated as second-order perturbations.

The {\sc AutoStructure} code is used to compute energy levels, oscillator strengths, excitation and photoionization cross sections, as well as autoionization rates, among other properties. These can be calculated at different levels of resolution: configuration resolution (configuration average, CA), term resolution (LS coupling), or level resolution (intermediate coupling, IC), utilizing semi-relativistic, kappa-averaged wavefunctions. The code also allows for the inclusion of configuration mixing.

Finally, the {\sc fac} code uses a fully relativistic approach to solve the Dirac equation. Quantum electrodynamics effects, primarily arising from the Breit interaction, vacuum polarization, and electron self-energy, are included in the standard procedures of the code. Like the other programs mentioned, {\sc fac} has been extensively used for interpreting spectroscopic data in both laboratory and astrophysical contexts.

\section{Results}
\label{sec:results}
\subsection{Atomic Structure}
\label{subsec:atomic}

The ion $W^{5+}$ is isoelectronic to Thulium ($Z=69$), having a ground 
configuration $[$Xe$]$ $4f^{14}5d$.
This configuration has two levels separated by about 1 eV of energy. 
The lowest level of the 
ground configuration is $(4f^{14}5d)_\frac{3}{2}$ and the 
second level is $(4f^{14}5d)_\frac{5}{2}$. 
For most of the relevant electronic densities and temperatures, 
both of these two levels are significantly occupied and in particular, 
the population of the $J=\frac{5}{2}$ is higher than those of the 
ground level.
We found two radiative transitions well separated and distinguishable  
in the experimental spectra, corresponding to the 
$(4f^{14}5f)_\frac{5}{2} \rightarrow (4f^{14}5d)_\frac{3}{2}$
transition at $\lambda_1=382.13$ \AA, and 
$(4f^{14}5f)_\frac{7}{2} \rightarrow (4f^{14}5d)_\frac{5}{2}$
transition at $\lambda_1=394.07$ \AA.

We include in our calculations, the $4f^{14}5l ~(l=2-4)$, $4f^{14}6l ~ 
(l=0-4)$, and $4f^{14}7l ~ (l=0-4)$.
We also include configuration-interaction (CI) between those configurations 
and $4f^{13}5d^2$, $4f^{13}5d6s$, $4f^{13}5d6p$, $4f^{13}6s^2$, 
$4f^{13}6s6p$, and also with the open $5p$-shell $4f^{14}5p^{5}5d^2$, 
$4f^{14}5p^{5}5d6s$, $4f^{14}5p^{5}6s^2$, $4f^{14}5p^{5}5d6p$, 
and $4f^{14}5p^{5}6s6p$.
This structure conforms to a total of 436 levels, the 
model used in HULLAC calculations. 
Excluding the last $4f^{14}5p^{5}6s6p$ configuration, the structure 
has 430 levels, and this is the model used for the AS and FAC calculations.
We show in Table~\ref{table:structure} the detailed results 
obtained with the three different calculations only for the 
four levels involved in the transitions.

\begin{table}[h!]
\footnotesize
\caption{Atomic structure for the 4-levels of W$^{5+}$ 
involved in the 382.13 \AA~and the 394.07 \AA~transitions.}
\label{table:structure}
\begin{center}
\begin{tabular}{ccccc}
\br
 & \multicolumn{3}{c}{E (eV)} &  \\
\cline{2-4} 
Index & {\sc hullac} & {\sc as} & {\sc fac} &  Level  \\
\br 
1 & 0.0000 & 0.0000 & 0.0000 & $(4f^{14}5d)_{3/2}$   \\  
2 & 0.9718 & 1.3653 & 1.0142 & $(4f^{14}5d)_{5/2}$   \\ 
3 & 31.769 & 31.823 & 31.729 & $(4f^{14}5f)_{5/2}$   \\  
4 & 31.841 & 31.973 & 31.814 & $(4f^{14}5f)_{7/2}$   \\ 
\br 
\end{tabular}
\end{center}
\end{table}

\begin{center}
\begin{figure}[h!]
\includegraphics[width=0.95\textwidth]{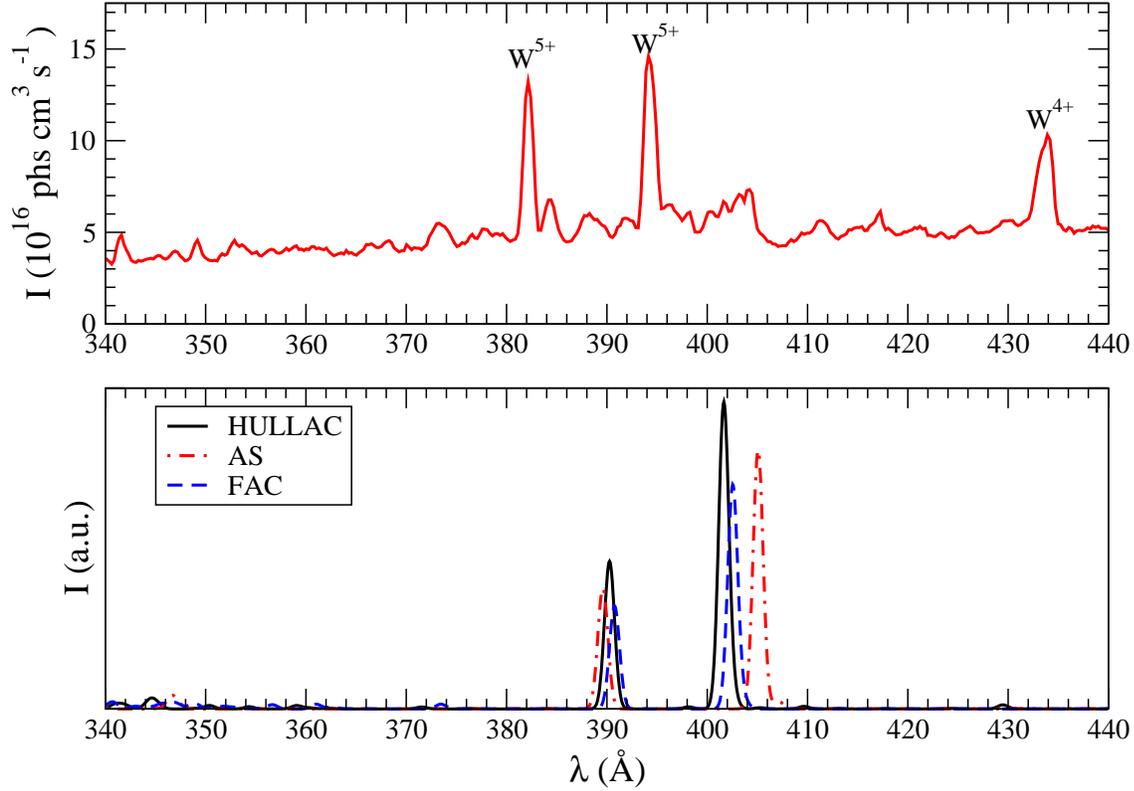}
\caption{EAST spectrum and theoretical collisional-radiative 
synthetic spectra for the $5f \rightarrow 5d$ transitions 
of the W$^{5+}$ ion. 
Upper graph: EAST Experimental spectra (\#100300, at $t=5.200$).
Lower graph: synthetic spectra calculated with {\sc HULLAC},
{\sc AS}, and {\sc FAC}.
}
\label{fig:spectra}
\end{figure}
\end{center}

Figure \ref{fig:spectra} shows the spectra observed from 
EAST discharge (\#100300) at a time $t=5.200$ s, corresponding 
to the maximum of the W burst. 
The figure also includes our three independent W$^{5+}$ synthetic 
spectra, which were obtained by solving the fully collisional-radiative model.

As the figure shows, the agreement between the three theoretical 
calculations is satisfactory, although the results are consistently higher 
than the experimental values by about 7 \AA. 
Surprisingly, another simpler calculation, including only 20 levels 
from the $4f^{14}5p^65l ~(l=2-4)$, $4f^{14}5p^66l ~(l=0-3)$, and 
$4f^{14}5p^67l ~(l=0-3)$ produces transition energies much closer than the 
experimental values, above the experimental for only 2 \AA, in all 
three cases. 
However, these calculations do not take into account the configuration 
mixings between the  $4f^{14}5f$ and the 
$4f^{13}5d^2$ configurations, which account for a total mixing of 
about 11\% for both the $J=5/2$ and $J=7/2$ levels.
This indicates that the latter model cannot be considered valid, 
and some serendipity is being introduced, 
which deserves to be investigated in the future.

\clearpage
\newpage
\subsection{Excitation and Radiative Rates}

The three codes have been employed for the calculation of the 
electron-impact excitation rate coefficients. 
The three codes use the distorted-waves approximation, 
which has been quite successful in the determination of excitation 
cross-sections for highly ionized species, where the effects of 
correlations are usually not significant. 
For W$^{5+}$, the coupling in the continuum is sufficiently small 
to allow the use of this approximation with reasonable accuracy.
The overall agreement between the three codes is about $40\%$, 
for the range $20 < T_e < 100$ eV.

We also have calculated the radiative transition rate coefficients 
with the three codes, obtaining values that agree $20\%$ for 
the dipole-allowed transitions. 
The values of the radiative transitions and the electron-impact 
excitation rate coefficients for the four levels involved in the 
$5f \rightarrow 5d$ lines are listed in Table~\ref{table:rates}, 
for {\sc hullac}, {\sc autostructure}, and {\sc fac} calculations.

\begin{table}[b]
\footnotesize
\centering
\caption{Radiative transitions $A$ and electron-impact 
excitation rate coefficients (in cm$^3$s$^{-1}$) for the 4 levels 
involved in the W$^{5+}$ lines. 
Note that aE$\pm$b means a $\times 10^{\pm b}$.}
\label{table:rates}
\begin{tabular}{ccccccccc}
\br
\multicolumn{9}{c}{{\sc hullac}} \\
 & & & \multicolumn{6}{c}{T$_e$ (eV)} \\
\cline{4-9}
Lower & Upper & A (s$^{-1}$) & 20 & 40 & 60 & 80 & 90 & 100 \\
\hline
1 & 2 & 5.20E+00 & 3.78E-09 & 2.47E-09 & 1.89E-09 & 1.56E-09 & 1.45E-09 & 1.35E-09 \\
1 & 3 & 1.61E+10 & 3.36E-09 & 5.88E-09 & 6.84E-09 & 7.27E-09 & 7.40E-09 & 7.49E-09 \\
2 & 3 & 1.16E+09 & 4.54E-10 & 6.56E-10 & 6.80E-10 & 6.68E-10 & 6.59E-10 & 6.49E-10 \\
1 & 4 & 1.01E+01 & 3.29E-10 & 4.27E-10 & 4.01E-10 & 3.62E-10 & 3.44E-10 & 3.27E-10 \\
2 & 4 & 1.83E+10 & 3.92E-09 & 6.66E-09 & 7.65E-09 & 8.09E-09 & 8.21E-09 & 8.30E-09 \\
3 & 4 & 2.18E-03 & 4.91E-09 & 3.15E-09 & 2.42E-09 & 2.02E-09 & 1.87E-09 & 1.75E-09 \\
\br
\\
\multicolumn{9}{c}{{\sc autostructure}} \\ 
 & & & \multicolumn{6}{c}{T$_e$ (eV)} \\
\cline{4-9}
Lower & Upper & A (s$^{-1}$) & 20 & 40 & 60 & 80 & 90 & 100 \\
\br
1  &  2    & 1.44E+01 & 8.62E-09 & 4.81E-09 & 3.36E-09 & 2.61E-09 & 2.35E-09 & 2.15E-09 \\
1  &  3    & 1.64E+10 & 2.74E-09 & 4.94E-09 & 5.89E-09 & 6.37E-09 & 6.52E-09 & 6.63E-09 \\
2  &  3    & 1.08E+09 & 4.50E-10 & 6.14E-10 & 6.27E-10 & 6.16E-10 & 6.08E-10 & 6.00E-10 \\
1  &  4    & 1.04E+01 & 3.63E-10 & 4.36E-10 & 3.96E-10 & 3.53E-10 & 3.34E-10 & 3.17E-10 \\
2  &  4    & 1.71E+10 & 3.34E-09 & 5.76E-09 & 6.74E-09 & 7.22E-09 & 7.36E-09 & 7.47E-09 \\
3  &  4    & 2.02E-02 & 1.01E-08 & 5.46E-09 & 3.79E-09 & 2.95E-09 & 2.67E-09 & 2.44E-09 \\
\br
\\
\multicolumn{9}{c}{{\sc fac}} \\
 & & & \multicolumn{6}{c}{T$_e$ (eV)} \\
\cline{4-9}
Lower & Upper & A (s$^{-1}$) & 20 & 40 & 60 & 80 & 90 & 100 \\
\hline
1  & 2    & 5.90E+00 & 5.07E-09 & 3.19E-09 & 2.40E-09 & 1.96E-09 & 1.81E-09 & 1.68E-09 \\
1  & 3    & 1.28E+10 & 2.41E-09 & 4.27E-09 & 5.00E-09 & 5.35E-09 & 5.46E-09 & 5.54E-09 \\
2  & 3    & 9.40E+08 & 4.31E-10 & 6.04E-10 & 6.16E-10 & 5.98E-10 & 5.87E-10 & 5.76E-10 \\
1  & 4    & 9.29E+00 & 3.39E-10 & 4.31E-10 & 3.99E-10 & 3.56E-10 & 3.36E-10 & 3.18E-10 \\
2  & 4    & 1.50E+10 & 3.11E-09 & 5.33E-09 & 6.16E-09 & 6.53E-09 & 6.64E-09 & 6.72E-09 \\
3  & 4    & 3.72E-03 & 5.70E-09 & 3.39E-09 & 2.57E-09 & 2.14E-09 & 1.99E-09 & 1.86E-09 \\
\br	 
\end{tabular}
\end{table}

\pagebreak
\clearpage
\newpage
\subsection{Electron-Impact Ionization Rates}
\label{subsec:ionization}

For the calculation of the electron-impact ionization rate 
coefficients, we used the distorted-waves approximation.
However, even by using this relatively simple approach, the 
calculations are cumbersome. 
First, the total direct ionization cross section $\sigma^{DI}$ 
from the ground configuration levels 
involves the sum of the contribution of the ionization 
of the 4$f$, 5$s$, 5$p$, and 5$d$ subshells-electrons:
\begin{equation}
e^{-}+[Kr] 4d^{10} 5s^{2} 5p^{6} 4f^{14} 5d \rightarrow[Kr]4d^{10} 
\left\{
   \begin{array}{c}
    5s^{2} 5p^{6} 4f^{14} \\
    5s^{2} 5p^{6} 4f^{13} 5d \\
    5s^{2} 5p^{5} 4f^{14} 5d \\
    5s 5p^{6} 4f^{14} 5d
    \end{array}
\right\}+2 e^{-}
\end{equation}

Second, the indirect pathway to ionization, through inner-shell 
excited intermediate levels must be taken into consideration since 
the cross-sections of these processes are much higher than the 
corresponding to the direct ionization, as indicated in 
the experimental ionization measurements by Stenke
{\it et al.} \cite{Stenke:95} and by Spruck {\it et al.} \cite{Spruck:14}.
The total cross-section $\sigma^{EA}_C$ for excitation-autoionization 
from an initial level $g$, to any final level $k$ of the
following ion, through inner-shell excitation 
to any intermediate autoionizing level $j$ within a 
given configuration or complex $C$ is given by \cite{Mitnik:97}
\begin{eqnarray}
\sigma^{EA}_C(E) &=& \sum_{j \in C} \sigma_{gj}(E) 
\left[
\frac{
\sum_k A^a_{jk} + \sum_i A_{ji} B^a_i
}
{
\sum_k A^a_{jk} + \sum_i A_{ji}
}
\right] \equiv 
\sum_{j \in C} \sigma_{gj}(E)  \, B^a_j
\label{eq:EA}
\end{eqnarray}
where $\sigma_{gj}$(E) is the cross section for electron-impact 
excitation from $g$ to $j$ as a function of the incident 
electron kinetic energy $E$. 
$A^a_{jk}$ is the rate coefficient for autoionization from 
$j$ to $k$, and $A_{ji}$ is the Einstein coefficient for 
spontaneous emission from $j$ to any lower-lying level $i$.
$B^a_j$ is the multiple branching ratio for autoionization 
from level $j$, defined by the bracket term. 
This term contains, in turn, the effective branching ratio 
$B^a_i$ for further autoionization from level $i$, defined 
by a similar recursive expression \cite{Mitnik:97}. 
The total excitation-autoionization cross-section is given by 
\begin{equation}
\sigma^{EA}(E) = \sum_C \sigma^{EA}_C(E) \, .
\end{equation}

In our calculation, we include excitations from the 
$4f$, $5p$, and $5s$ subshells of the ground configuration. 
These excitations are graphically represented as follows:
\begin{equation}
e^{-}+[Kr] 4d^{10} 5s^{2} 5p^{6} 4f^{14} 5d \rightarrow[Kr]4d^{10} 
\left\{
   \begin{array}{c}
    5s^{2} 5p^{6} 4f^{13} 5d nl \\
    5s^{2} 5p^{5} 4f^{14} 5d nl \\
    5s 5p^{6} 4f^{14} 5d nl 
    \end{array}
\right\}+ e^{-}
\end{equation}
where we take all the possibles $5 \leq n \leq 15$ and $l \leq 4$.

In a low-ionized species such as W$^{5+}$, the radiative rates
from the doubly excited levels are quite small compared
to the autoionizing rates. 
Therefore, we can neglect the multiple branching ratios in 
(\ref{eq:EA}). 
In that way, we assume that once the excited
electron reaches an autoionizing level, it autoionizes and does 
not follow further cascades.

The total electron-impact single ionization 
cross-section is
\begin{equation}
\sigma^{EIA}(E)= \sigma^{DI}(E) + \sigma^{EA}_C(E) \, .
 \label{eq:totalS}
\end{equation}

\begin{center}
\begin{figure}[h!]
\centering\includegraphics[width=0.49\textwidth]{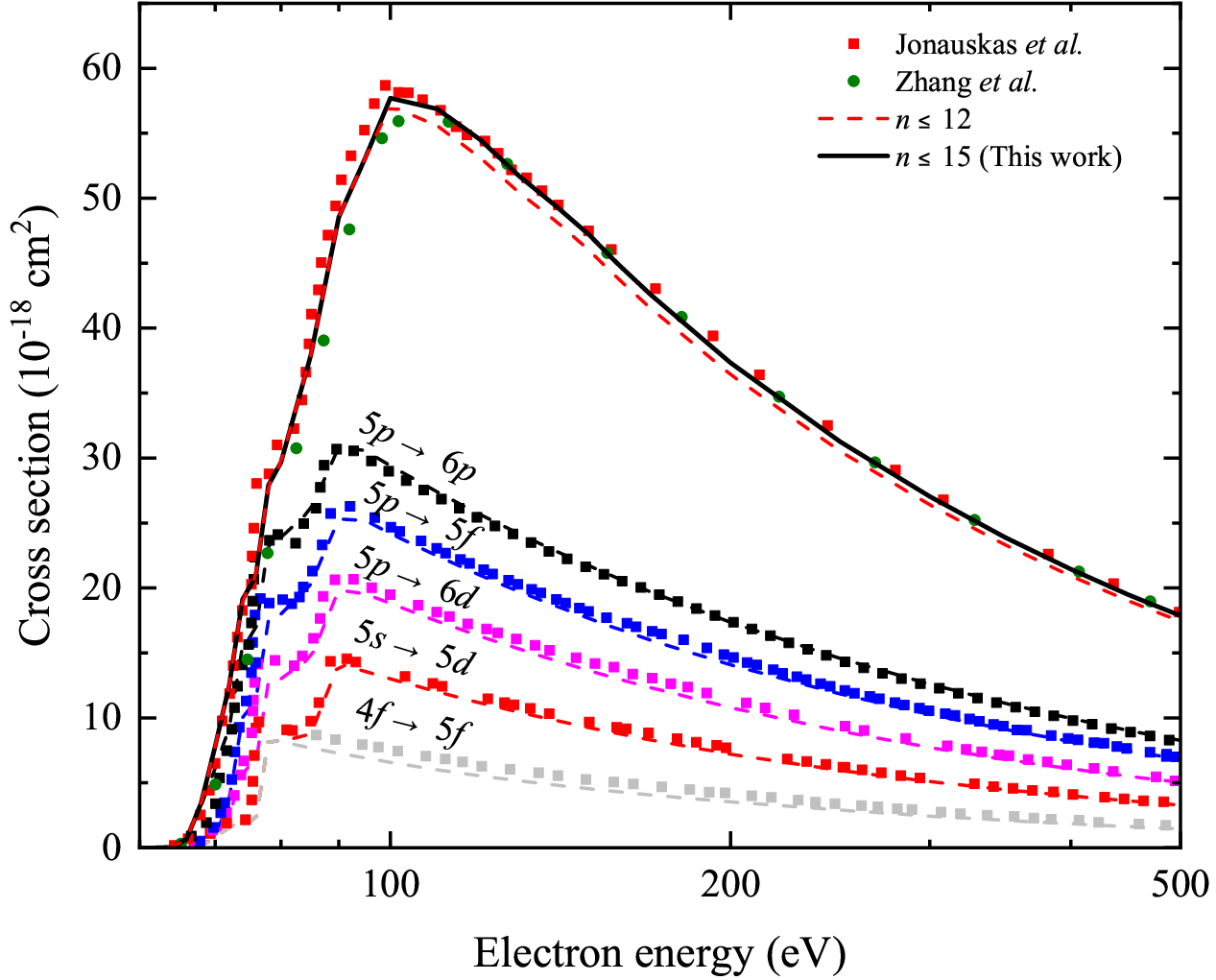}
\includegraphics[width=0.49\textwidth]{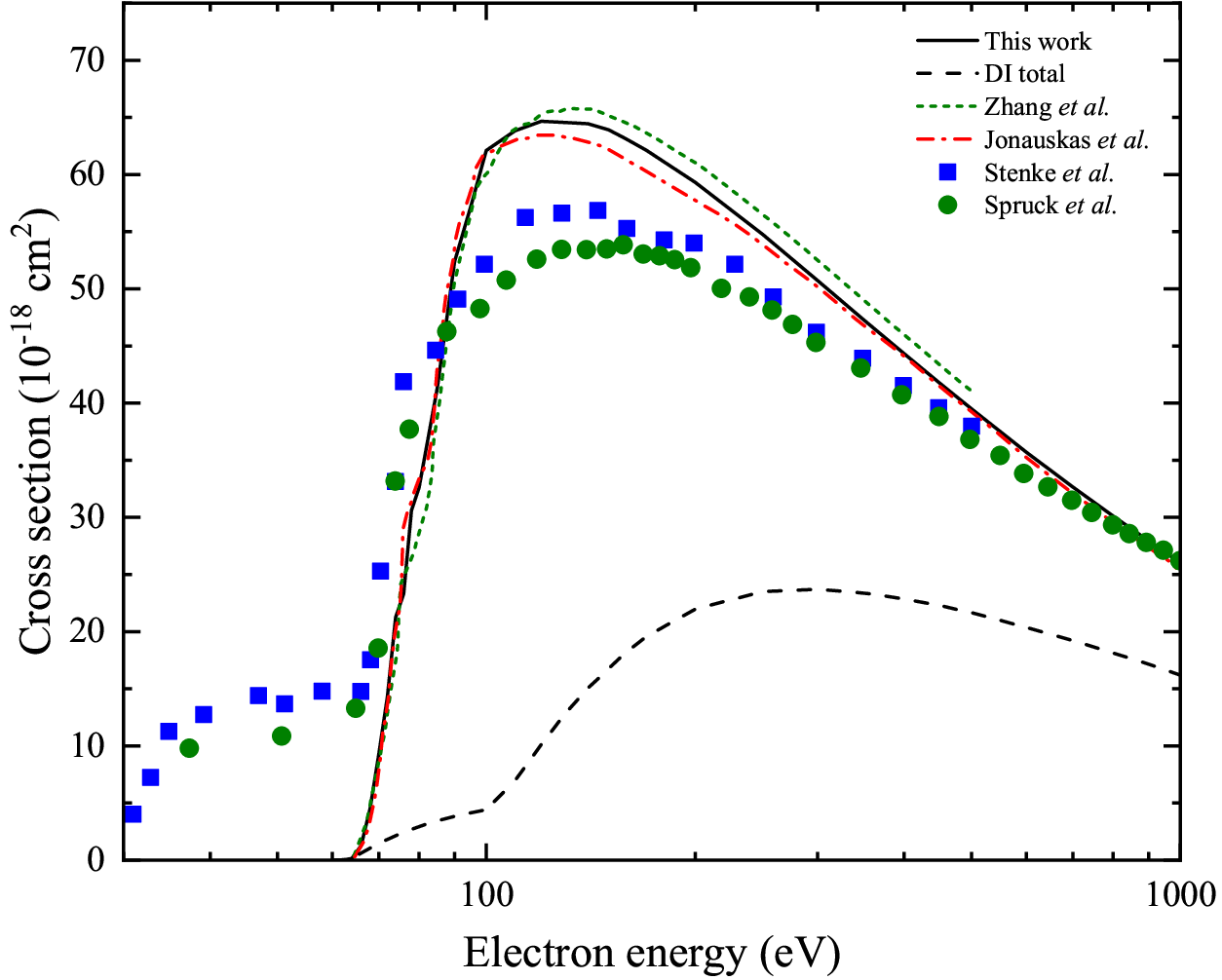}
\caption{
Left: Excitation-Autoionization cross-sections  
from the ground state of $W^{5+}$. 
Color-dashed lines: Present calculations of the accumulated 
partial EA cross sections $\sigma^{EA}_C$ for the principal 
manifolds $C$. 
Color squares: Theoretical results from Jonauskas {\it et al.}  \cite{Jonauskas:19}. 
Red circles: Total theoretical results by 
Zhang {\it et al.} \cite{ZhangIoniz:18}.
Black-solid lines: Present calculations for total $\sigma^{EA}$, 
considering $n \leq 15$.

Right: Total ionization cross-sections from the ground level.
Black solid line: Present calculations of the 
total ionization cross-section $\sigma^{EIA}$
(Eq.~\ref{eq:totalS}).
Black-dashed line: Total direct-ionization cross-section $\sigma^{DI}$. 
Green-dashed line: Theoretical results from Jonauskas {\it et al.}  \cite{Jonauskas:19}. 
Red-dot dashed line: Calculations by Zhang {\it et al.} \cite{ZhangIoniz:18}. 
Blue squares: Experimental results of Stenke {\it et al.} 
\cite{Stenke:95}; 
Green circles: Experimental results of Spruck {\it et al.} \cite{Spruck:14}. 
}
\label{fig:ionization}
\end{figure}
\end{center}

Since the calculations are demanding, we only use one computational 
code for the ionization cross-section results. 
Also, for comparisons with other calculations, we choose to 
use the {\sc fac} code. 
In Fig.~\ref{fig:ionization}, we show the contribution of the 
different channels to the electron-impact ionization cross-sections 
of $W^{5+}$. 
In the left part of the figure, we concentrate on the indirect 
excitation-autoionization channels $\sigma^{EA}$. 
The figure shows the accumulated contribution from the essential  
manifolds (dashed-color curves), together with the theoretical 
values reported by Jonauskas {\it et al.} \cite{Jonauskas:19}. 
The results are in excellent agreement and the slight differences 
can be attributed to the lack of some configuration interactions 
that we have not included in the calculations.
On the other hand, Jonauskas {\it et al.} calculated the EA contribution 
up to the $n=12$ manifold, whereas we also included the $n=(14,15)$ 
configurations.

In the right part of the figure, the total ionization cross-sections 
$\sigma^{EIA}$ are displayed. 
We intended to show the individual contribution of the 
ionization from the ground level $(4f^{14}5d)_\frac{3}{2}$ and 
from the metastable  $(4f^{14}5d)_\frac{5}{2}$, but both 
cross-sections are very similar and are not easily distinguishable 
in the graph. 
The figure shows our results, together with the theoretical 
calculations reported by Jonauskas {\it et al.}  \cite{Jonauskas:19} and 
also the results given by Zhang {\it et al.} \cite{ZhangIoniz:18}. 
As expected, the agreement is excellent. The differences can 
be attributed to the configurations used in the potential optimization  
and the configurations considered for the mixing interactions.
To be able to assess the importance of the indirect channels, 
we include in the figure the total direct ionization $\sigma^{DI}$  
in dashed lines.
The figure also shows the experimental results given by 
Stenke {\it et al.} \cite{Stenke:95}, and more recent 
calculations from Spruck {\it et al.} \cite{Spruck:14}. 
The distorted-waves are known for generally overestimation of the 
cross-sections. Both experimental results show a high presence of
metastable contributions, so the discrepancies between theoretical 
and experimental results could be even more significant.

\begin{table}[b]
\footnotesize
\centering
\caption{Total ionization rate coefficients from the metastable 
levels of W$^{5+}$, in cm$^3$s$^{-1}$. 
Note that aE-b means a $\times 10^{-b}$.}
\label{table:totalS}
\begin{tabular}{ccccccc}
\br
 &   \multicolumn{6}{c}{T$_e$ (eV)} \\
\cline{2-7}
Metastable & 20 & 40 & 60 & 80 & 90 & 100 \\
\br                
1 & 2.01E-09 & 1.15E-08 & 2.04E-08 & 2.70E-08 & 2.96E-08 & 3.18E-08 \\
2 & 2.07E-09 & 1.18E-08 & 2.08E-08 & 2.75E-08 & 3.01E-08 & 3.23E-08 \\	
\br               
\end{tabular}
\end{table}

\pagebreak
\clearpage
\newpage
\subsection{S/XB results}
\subsubsection{Simple Model (4-levels) } 
~ \\

We found that a simple model having only four levels, namely, 
the ground level $(4f^{14}5d)_{3/2}$, the 
metastable $(4f^{14}5d)_{5/2}$, and the 
two upper-excited levels, $(4f^{14}5f)_{5/2}$ and $(4f^{14}5f)_{7/2}$, 
can be used to explain qualitatively many of the basic 
features in the determination of impurity flux.
A detailed derivation of the S/XB partial coefficients is given in 
\ref{appendix:SXB}. 
The final expressions for these coefficients are given in 
Eq.~(\ref{eq:SXB4}).

\begin{center}
\begin{figure}[h]
\centering\includegraphics[width=0.75\textwidth]{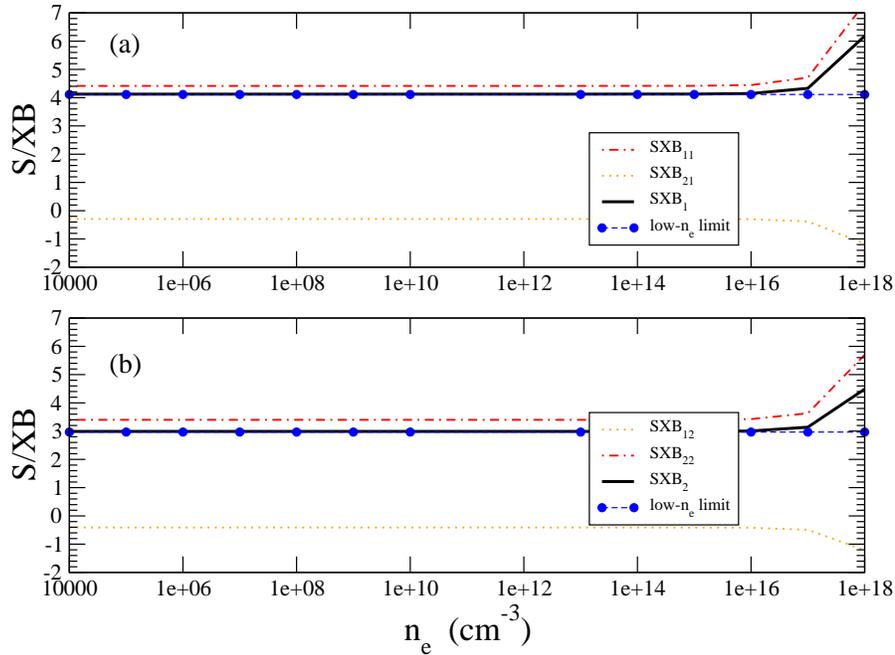}
\caption{S/XB coefficients, according to the definition in 
Eq.~(\ref{eq:sxbn}), for a 4-level model. 
(a) S/XB for the $3 \rightarrow 1$ transition 
($(4f^{14}5f)_{5/2} \rightarrow (4f^{14}5d)_{3/2}$). 
(b) $4 \rightarrow 2$ transition 
($(4f^{14}5f)_{7/2} \rightarrow (4f^{14}5d)_{5/2}$), in W$^{5+}$.
The electronic temperature is $T_e=60$ eV. 
The low-density limit curves correspond to the approximation given 
in Eq.~(\ref{eq:lowdens}). 
}
\label{fig:lowdens}
\end{figure}
\end{center}

We show, in Figure~\ref{fig:lowdens}, the S/XB coefficients corresponding 
to the transition $3 \rightarrow 1$ (upper part (a)), and in the 
lower part (b), the coefficients corresponding to the $4 \rightarrow 2$ 
transition. 
The coefficients for this table have been calculated using the 
rates obtained with {\sc fac} code for a typical electronic temperature 
$T_e=60$ eV.
As is expected, due to the high population of the metastable, the leading 
terms are the diagonal ${\cal SXB}_{11}$ and ${\cal SXB}_{22}$, 
being the cross terms a small negative contribution.

We can perform a further approximation, focusing on the low-density 
region, in order to understand the overall behavior of the S/XB coefficients. 
In this approximation, we neglect first the collisional decays 
that are proportional to the electron density and compete with 
stronger radiative decays. 
We also make use of the rates listed in Table \ref{table:rates}, 
neglecting the relatively small terms, for example, 
$(A_{41} + A_{42} + A_{43}) \approx A_{42}$. 
Under these assumptions, the low-density limit expressions 
for the S/XB coefficients are
\begin{eqnarray}
{\cal SXB}_{11} & \approx &  \frac{ S_1 }{ Q_{13} \, \frac{A_{31}}{A_{31}+A_{32}} }
\nonumber \\
{\cal SXB}_{21} & \approx & - \frac{ S_2 }{ Q_{24} \, \frac{Q_{13}}{Q_{14}} }
\nonumber \\
{\cal SXB}_{12} & \approx &  -\frac{ S_1 }{ Q_{13} \, \frac{Q_{24}}{Q_{23}} }
\nonumber \\
{\cal SXB}_{22} & \approx &  \frac{ S_2 }{ Q_{24} }
\label{eq:lowdens}
\end{eqnarray}
which are all density independent. 
As is shown in Figure~\ref{fig:lowdens}, the approximated S/XB results 
are in excellent agreement with the exact expressions given in (\ref{eq:SXB4}). 
The results of this approximation not only match the exact results at 
low electronic densities but this agreement is still maintained for 
densities that are above the standard operating values, 
even up to values such as $n_e \approx 10^{17}$ cm$^{-3}$.

\begin{center}
\begin{figure}[h]
\centering\includegraphics[width=0.75\textwidth]{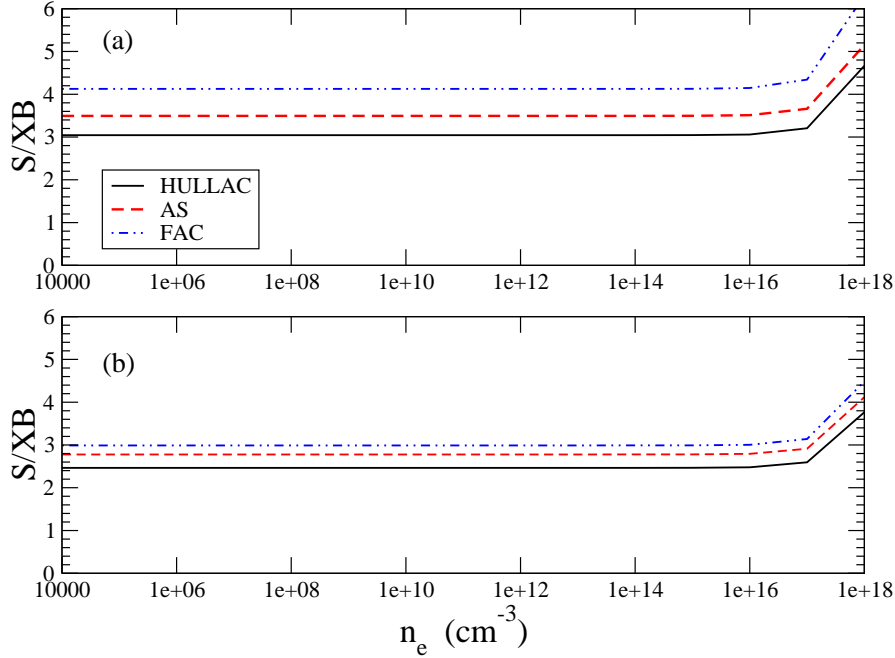}
\caption{Comparisons of the S/XB coefficients obtained with 
the three independent calculations using {\sc hullac}, {\sc as}, and 
{\sc fac} codes for the 4-level model. 
(a) S/XB for the $3 \rightarrow 1$ transition 
($(4f^{14}5f)_{5/2} \rightarrow (4f^{14}5d)_{3/2}$). 
(b) $4 \rightarrow 2$ transition 
($(4f^{14}5f)_{7/2} \rightarrow (4f^{14}5d)_{5/2}$), in W$^{5+}$.
The electronic temperature is $T_e=60$ eV. 
}
\label{fig:4levsxbte3}
\end{figure}
\end{center}

We made comparisons between the S/XB coefficients resulting from the three independent 
calculations from {\sc hullac}, {\sc as}, and {\sc fac} computational 
codes. 
The results are displayed in Figure~\ref{fig:4levsxbte3} for an 
electronic temperature $T_e=60$ eV, and as before, the upper part 
shows the results for the $3 \rightarrow 1$ transition, and the 
lower part $4 \rightarrow 2$.

\begin{center}
\begin{figure}[h]
\centering\includegraphics[width=0.75\textwidth]{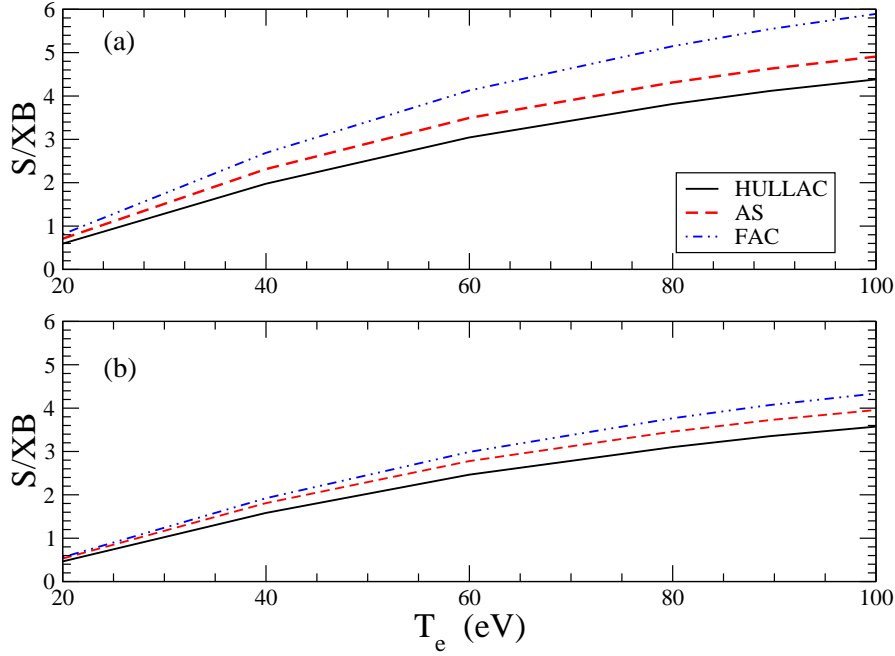}
\caption{Comparisons of the S/XB coefficients obtained 
for the 4-level model. 
(a) S/XB for the $3 \rightarrow 1$ transition 
($(4f^{14}5f)_{5/2} \rightarrow (4f^{14}5d)_{3/2}$). 
(b) $4 \rightarrow 2$ transition 
($(4f^{14}5f)_{7/2} \rightarrow (4f^{14}5d)_{5/2}$), in W$^{5+}$.
The electronic density is $n_e=10^{13}$ cm$^{-3}$. 
}
\label{fig:4levsxbne3}
\end{figure}
\end{center}

The same results are shown in Figure~\ref{fig:4levsxbne3} for the 
electronic density $n_e=10^{13}$ cm$^{-3}$, which is within the normal 
range of operation in the EAST plasmas.
The discrepancies between the different calculations are consistent 
with the differences obtained for the values of the rates shown in 
Table~\ref{table:rates}, in particular, for the $A_{31}$ 
radiative decay rates.

\newpage
\subsubsection{Full Model (430-levels)}
  ~ \\

\begin{center}
\begin{figure}[h]
\centering\includegraphics[width=0.75\textwidth]{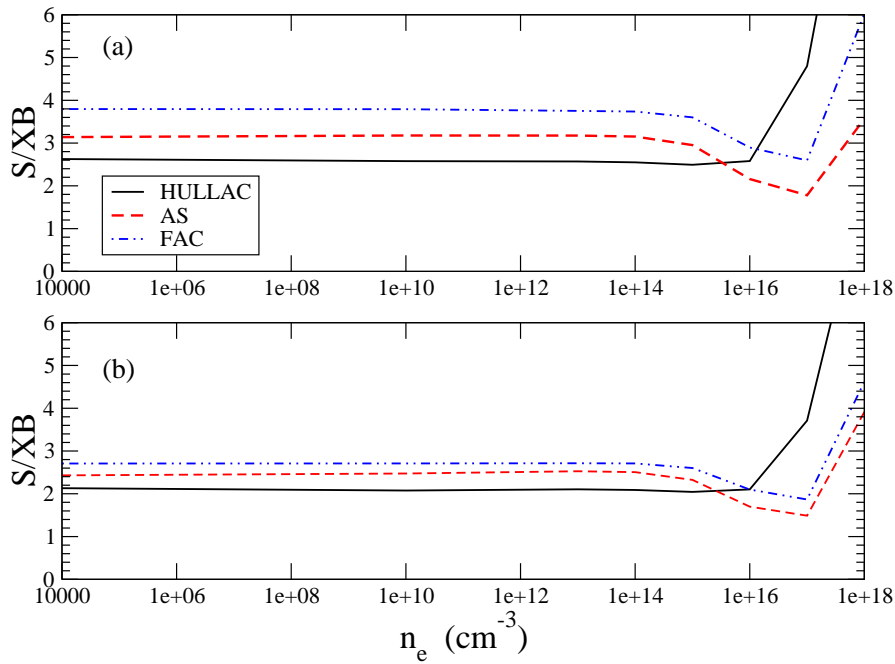}
\caption{Comparisons of the S/XB coefficients obtained with 
the three independent calculations using {\sc hullac}, {\sc as}, and 
{\sc fac} codes, for the 430-level model, in W$^{5+}$.
(a) S/XB for the $(4f^{14}5f)_{5/2} \rightarrow (4f^{14}5d)_{3/2}$ transition.  
(b) $(4f^{14}5f)_{7/2} \rightarrow (4f^{14}5d)_{5/2}$ transition. 
The electronic temperature is $T_e=60$ eV. 
}
\label{fig:sxbte3}
\end{figure}
\end{center}

\begin{center}
\begin{figure}[h]
\centering\includegraphics[width=0.75\textwidth]{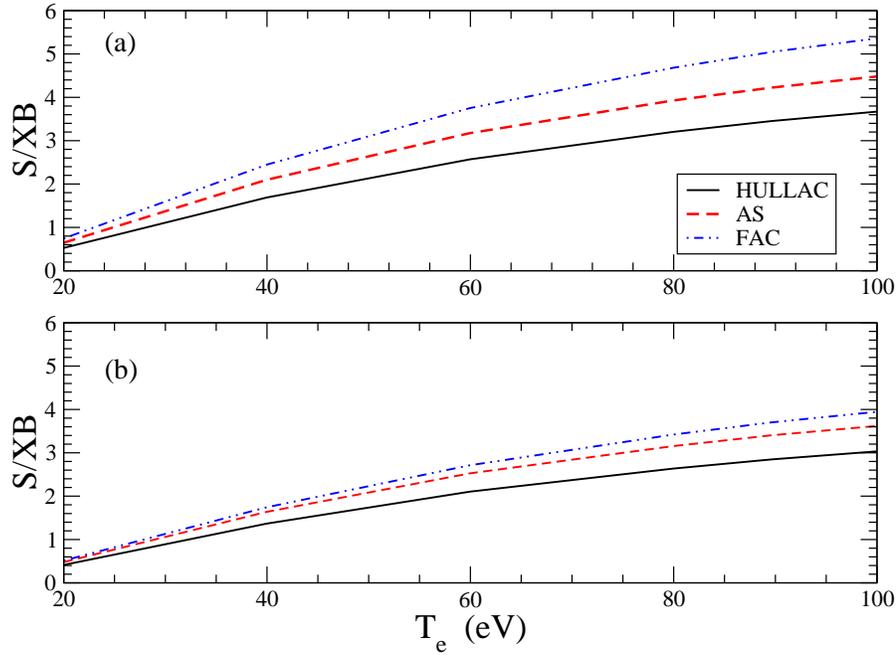}
\caption{Comparisons of the S/XB coefficients obtained 
for the 430-level model, in W$^{5+}$. 
(a) S/XB for the $(4f^{14}5f)_{5/2} \rightarrow (4f^{14}5d)_{3/2}$ transition.  
(b) $(4f^{14}5f)_{7/2} \rightarrow (4f^{14}5d)_{5/2}$ transition. 
The electronic density is $n_e=10^{13}$ cm$^{-3}$. 
}
\label{fig:sxbne3}
\end{figure}
\end{center}

We included all the configurations listed in Sec.~\ref{subsec:atomic}  
and solved the corresponding collisional-radiative equations, 
obtaining the S/XB ratios for the same transitions studied previously.
The results are displayed in Fig.~\ref{fig:sxbte3} for a fixed 
electronic temperature $T_e=60$ eV and in Fig.~\ref{fig:sxbne3}, 
for a fixed electronic density $n_e=10^{13}$ cm$^{-3}$. 
We note that for both lines, the S/XB ratios appear 
nearly constant around an electron density from 
10$^{4}$ cm$^{-3}$, to a density of approximately $10^{14}$ 
cm$^{-3}$. 
This justifies the previous approach in which we considered only four 
levels and the possibility of using the low-density approximation 
to obtain a simple and quick determination of the incoming flux. 
At densities around $10^{15}$ cm$^{-3}$ the S/XB becomes sensitive to 
the collisional-radiative solutions. 
Therefore, we found substantial 
differences between the calculations, particularly  
between {\sc hullac} results and the other two.
Beyond $n_e > 10^{17}$ cm$^{-3}$, the plasma reaches the local 
thermodynamic equilibrium density regime, and then the 
S/XB ratios increase linearly with the density.

The detailed results for electronic densities in the range 
$10^4$ cm$^{-3}  < n_e < 10^{18}$ cm$^{-3}$ and electronic 
temperatures in the range $20 < T_e < 100$ eV are presented 
in Tables \ref{table:sxb430hul}, \ref{table:sxb430as}, 
and \ref{table:sxb430fac} for the calculations realized with 
{\sc hullac}, {\sc autostructure}, and {\sc fac}, respectively.
The agreement between the calculations for the 
$(4f^{14}5f)_{5/2} \rightarrow (4f^{14}5d)_{3/2}$ transition 
(the $3 \rightarrow 1$ transition in the 4-level model) is 
of about $40\%$ for $T_e=50$ eV and 
densities up to $n_e=10^{15}$ cm$^{-3}$ in 
the worst case ({\sc hullac} vs. {\sc fac}). 
Fig.~\ref{fig:sxbne3} shows that the agreement is much better at lower 
temperatures. 
For the $(4f^{14}5f)_{7/2} \rightarrow (4f^{14}5d)_{5/2}$ transition, 
the agreement at $T_e=50$ eV is about $25\%$, it is much 
better for lower temperatures. 

\begin{table}[h!]
\footnotesize
\centering
\caption{SXB of W$^{5+}$ for the radiative transitions 
(Line 1) $(4f^{14}5f)_{5/2} \rightarrow (4f^{14}5d)_{3/2}$, and 
(Line 2) $(4f^{14}5f)_{7/2} \rightarrow (4f^{14}5d)_{5/2}$, from 
a 436-level collisional-radiative model, calculated 
with the HULLAC computational suite.}
\label{table:sxb430hul}
\begin{tabular}{cccccccccc}
\br
 & & \multicolumn{8}{c}{Electron density (cm$^{-3}$)} \\
\cline{3-10}			
 Line & T$_e$(eV)  &  10$^{4}$ & 10$^{10}$ & 10$^{13}$ & 10$^{14}$ & 
                   10$^{15}$ & 10$^{16}$ & 10$^{17}$ & 10$^{18}$  \\
\br 
1 & 20  & 0.543 & 0.532 & 0.525 & 0.519 & 0.502 & 0.504 & 0.804 & 1.710 \\
& 40  & 1.733 & 1.698 & 1.690 & 1.674 & 1.629 & 1.677 & 3.039 & 7.143 \\
& 60  & 2.626 & 2.577 & 2.569 & 2.549 & 2.492 & 2.579 & 4.798 & 11.90 \\
& 80  & 3.266 & 3.209 & 3.203 & 3.180 & 3.119 & 3.235 & 6.059 & 15.52 \\
& 90  & 3.516 & 3.458 & 3.452 & 3.429 & 3.367 & 3.494 & 6.543 & 16.92 \\
& 100 & 3.736 & 3.676 & 3.670 & 3.648 & 3.586 & 3.722 & 6.966 & 18.20 \\
\br 		 
2 & 20  & 0.425 & 0.414 & 0.412 & 0.408 & 0.393 & 0.384 & 0.570 & 1.470 \\
& 40  & 1.391 & 1.354 & 1.367 & 1.355 & 1.318 & 1.331 & 2.258 & 5.758 \\
& 60  & 2.130 & 2.076 & 2.105 & 2.089 & 2.045 & 2.103 & 3.712 & 9.203 \\
& 80  & 2.661 & 2.598 & 2.636 & 2.620 & 2.575 & 2.678 & 4.805 & 11.70 \\
& 90  & 2.873 & 2.807 & 2.848 & 2.833 & 2.790 & 2.912 & 5.250 & 12.72 \\
& 100 & 3.056 & 2.988 & 3.032 & 3.017 & 2.975 & 3.116 & 5.639 & 13.60 \\
\br
\end{tabular}
\end{table}

\begin{table}[h!]
\footnotesize
\centering
\caption{SXB of W$^{5+}$ for the radiative transitions 
(Line 1) $(4f^{14}5f)_{5/2} \rightarrow (4f^{14}5d)_{3/2}$, and 
(Line 2) $(4f^{14}5f)_{7/2} \rightarrow (4f^{14}5d)_{5/2}$, from  
a 430-level collisional-radiative model, calculated 
with the AS computational code.}
\label{table:sxb430as}
\begin{tabular}{cccccccccc}
\br
& & \multicolumn{8}{c}{Electron density (cm$^{-3}$)} \\
\cline{3-10}			
 Line & T$_e$(eV)  &  10$^{4}$ & 10$^{10}$ & 10$^{13}$ & 10$^{14}$ & 
                   10$^{15}$ & 10$^{16}$ & 10$^{17}$ & 10$^{18}$  \\
\br 
1 & 20  & 0.645 & 0.654 & 0.644 & 0.637 & 0.580 & 0.395 & 0.343 & 0.877 \\
& 40  & 2.076 & 2.107 & 2.099 & 2.080 & 1.921 & 1.352 & 1.136 & 2.457 \\
& 60  & 3.140 & 3.178 & 3.176 & 3.153 & 2.952 & 2.158 & 1.779 & 3.568 \\
& 80  & 3.889 & 3.927 & 3.928 & 3.905 & 3.694 & 2.787 & 2.272 & 4.356 \\
& 90  & 4.184 & 4.221 & 4.224 & 4.201 & 3.990 & 3.049 & 2.476 & 4.669 \\
& 100 & 4.442 & 4.476 & 4.480 & 4.457 & 4.249 & 3.285 & 2.657 & 4.929 \\
\br
2 & 20  & 0.469 & 0.480 & 0.479 & 0.474 & 0.430 & 0.293 & 0.249 & 0.674 \\
& 40  & 1.575 & 1.609 & 1.638 & 1.623 & 1.487 & 1.048 & 0.918 & 2.477 \\
& 60  & 2.431 & 2.474 & 2.526 & 2.507 & 2.326 & 1.699 & 1.487 & 3.913 \\
& 80  & 3.046 & 3.090 & 3.155 & 3.135 & 2.939 & 2.210 & 1.930 & 4.965 \\
& 90  & 3.293 & 3.336 & 3.404 & 3.385 & 3.186 & 2.425 & 2.115 & 5.390 \\
& 100 & 3.505 & 3.547 & 3.618 & 3.599 & 3.400 & 2.617 & 2.281 & 5.765 \\
\br
\end{tabular}
\end{table}

\begin{table}[h!]
\footnotesize
\centering
\caption{SXB of W$^{5+}$ for the radiative transitions 
(Line 1) $(4f^{14}5f)_{5/2} \rightarrow (4f^{14}5d)_{3/2}$, and 
(Line 2) $(4f^{14}5f)_{7/2} \rightarrow (4f^{14}5d)_{5/2}$, from  
a 430-level collisional-radiative model, calculated 
with the FAC computational code.}
\label{table:sxb430fac}
\begin{tabular}{cccccccccc}
\br
& & \multicolumn{8}{c}{Electron density (cm$^{-3}$)} \\
\cline{3-10}			
 Line & T$_e$(eV)  &  10$^{4}$ & 10$^{10}$ & 10$^{13}$ & 10$^{14}$ & 
                   10$^{15}$ & 10$^{16}$ & 10$^{17}$ & 10$^{18}$  \\
\br 
1 & 20  & 0.762 & 0.759 & 0.745 & 0.740 & 0.696 & 0.529 & 0.484 & 1.328 \\
& 40  & 2.484 & 2.479 & 2.448 & 2.436 & 2.326 & 1.828 & 1.670 & 4.094 \\
& 60  & 3.796 & 3.792 & 3.752 & 3.737 & 3.601 & 2.896 & 2.595 & 6.026 \\
& 80  & 4.730 & 4.725 & 4.681 & 4.666 & 4.525 & 3.710 & 3.273 & 7.351 \\
& 90  & 5.096 & 5.091 & 5.045 & 5.031 & 4.892 & 4.046 & 3.550 & 7.873 \\
& 100 & 5.414 & 5.409 & 5.361 & 5.348 & 5.211 & 4.343 & 3.790 & 8.314 \\
\br
2 & 20  & 0.520 & 0.518 & 0.512 & 0.510 & 0.482 & 0.371 & 0.323 & 0.830 \\
& 40  & 1.748 & 1.747 & 1.745 & 1.741 & 1.660 & 1.311 & 1.173 & 2.975 \\
& 60  & 2.707 & 2.709 & 2.714 & 2.710 & 2.603 & 2.098 & 1.866 & 4.619 \\
& 80  & 3.407 & 3.411 & 3.421 & 3.419 & 3.301 & 2.702 & 2.384 & 5.764 \\
& 90  & 3.685 & 3.690 & 3.702 & 3.701 & 3.580 & 2.951 & 2.595 & 6.203 \\
& 100 & 3.926 & 3.932 & 3.945 & 3.946 & 3.824 & 3.173 & 2.780 & 6.579 \\
\br	 
\end{tabular}
\end{table}

\newpage
\clearpage
\subsection{Tungsten Influx}

\begin{center}
\begin{figure}[h]
\centering\includegraphics[width=0.75\textwidth]{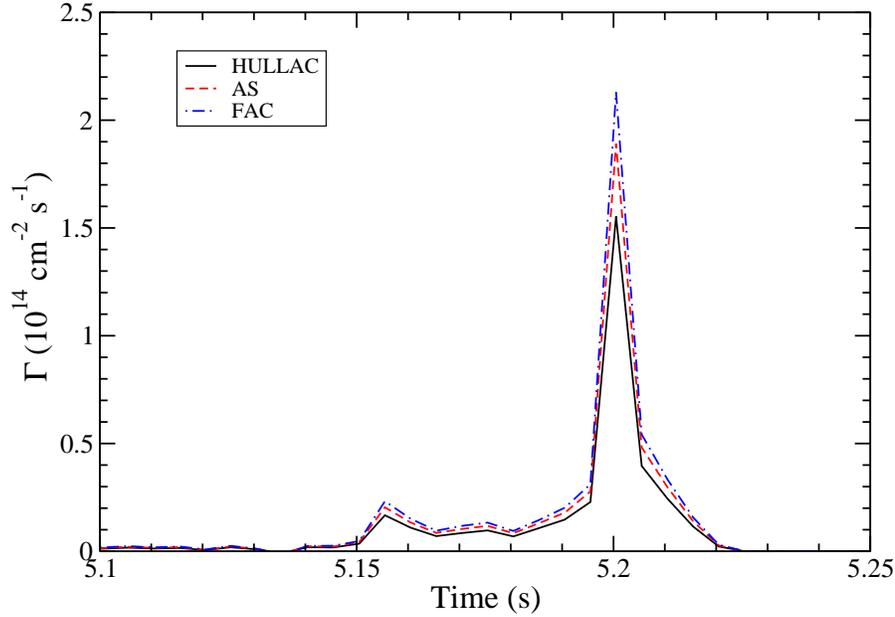}
\caption{Total tungsten influx from the plasma-facing surface, 
calculated for the discharge \#100300 at EAST, as a function of time. 
}
\label{fig:flux}
\end{figure}
\end{center}

The absolute calibration of the spectral line intensities 
\cite{ZhangLing:15,Dong:11,Dong:12,Lei:21} results in 
values of the $5f \rightarrow 5d$ transitions about 
$3 \times 10^{13}$ photons per cm$^{2}$ per second at 
the emission peak corresponding to the tungsten burst. 
Assuming that the tungsten ions are sputtered from a region in which 
the electronic temperature is $T_e=60$ eV and the electronic 
density is $n_e=10^{13}$ cm$^{-3}$, the 
S/XB ratios are of the order of 3, we can infer a total 
tungsten flux of the order of $10^{14}$ particles per 
cm$^{2}$ per second at the peak.
Indeed, we show in Fig.~\ref{fig:flux} the calculated 
total tungsten flux 
from the plasma-facing surface towards the interior of the 
EAST device, which follows the behavior of the line intensity 
in time for discharge \#100300, having a peak at $t=5.2$ seconds. 
The fluxes are calculated using three independent calculations: 
{\sc hullac}, {\sc as}, and {\sc fac}. 
The three calculatons agree at the peak between $45\%$ in the 
worst comparison case ({\sc hullac} vs. {\sc fac}).

\clearpage
\section{Conclusions}
\label{sec:conclusions}

In this study, we derive the relevant equations for calculating the S/XB coefficients in scenarios where, rather than assuming a single initial state, multiple metastable states with significant populations and influence on ionization processes are considered. Unlike excited states, metastable states primarily equilibrate their populations with those of neighboring ions. This allows for the expression of the relative populations of the excited states in terms of the populations of the metastable states, without the need for explicitly specifying the latter’s population values. Similarly, ionization events, which are dependent on the metastable populations, can be described in terms of the radiative emissions from the excited states.

Extensive theoretical calculations of atomic structure, radiative decay, 
electron-impact excitation, and electron-impact ionization have been carried 
out to generate the data necessary to model the $W^{5+}$ ion in a plasma environment. 
We use three different atomic computational codes -- {\sc fac}, {\sc hullac}, and {\sc autostructure} --
which allow us to assess the accuracy of our theoretical results.
For the electron-impact ionization, we included contributions from 
excitation-autoionization processes up to $n=15$ manifolds and 
calculated the rate coefficients for the total ionization from both the two 
$4f^{14}5d$ configuration levels.
 
Using these data, we solved a collisional-radiative model to obtain the effective population coefficients needed in to determine the S/XB ratios.
We selected the radiative transitions 
$(4f^{14}5f)_{5/2} \rightarrow (4f^{14}5d)_{3/2}$ and 
$(4f^{14}5f)_{7/2} \rightarrow (4f^{14}5d)_{5/2}$, which are 
prominent in the 320-460 \AA~range. 
The process of radiative cascade from energy terms higher than
the upper terms involved in the radiative transition was found to have a 
negligible effect on the S/XB ratios for both types of transitions at 
low and intermediate energy densities. 
Therefore, by using a simple four-level collisional-radiative model 
consisting of the two upper levels and the two lower metastables, 
we can generate S/XB ratios that agree very well with the 
430-level model. 
That is particularly useful because it allows us to derive simple analytical 
expressions that help us understand the principal mechanisms contributing 
to the ionization. 

\section{Acknowledgments}

This work was supported by the National Magnetic Confinement Fusion Energy R \& D Program of China (Grant No. 2022YFE03180400), National Natural Science Foundation of China (Grant Nos. 12322512,  12274352), and Chinese Academy of Sciences President's International Fellowship Initiative (PIFI) (Grant No. 2024PVA0074 ). 
DM acknowledges partial support from CONICET by Project No.
PIP11220200102421CO, and the ANPCyT by Project No. PICT-
2020-SERIE A-01931 in Argentina, and the Alliance of International Science Organizations (ANSO) Visiting Fellowship (ANSO-VF-2021-03), in China.
\appendix 
\section{Derivation of S/XB coefficients for 4-levels and 2 metastables}
\label{appendix:SXB}
\setcounter{section}{1}

The collisional-radiative equations (\ref{eq:crmatrix}) for 4-levels 
is:
\begin{eqnarray}
\left(
\begin{array}{c}
\frac{dn_1}{dt} \\
\frac{dn_2}{dt} \\
0 \\
0 \\
\end{array}
\right)
=
\left(
\begin{array}{cccc}
C_{11}  & C_{12} & C_{13} &  C_{14} \\
C_{21}  & C_{22} & C_{23} &  C_{24} \\
C_{31}  & C_{32} & C_{33} &  C_{34} \\
C_{41}  & C_{42} & C_{43} &  C_{44} 
\end{array}
\right)
\left(
\begin{array}{c}
n_1 \\
n_2 \\
n_3 \\
n_4
\end{array}
\right) \,\, ,
\hfill
\label{eq:Apopwithmeta}
\end{eqnarray}

~ \\
and for the excited-levels part of the matrix, the corresponding 
system of equations is
\begin{eqnarray}
\left(
\begin{array}{c}
C_{31} n_1 +  C_{32} n_2 +  C_{33} n_3 + C_{34} n_4 \\
C_{41} n_1 +  C_{42} n_2 +  C_{43} n_3 + C_{44} n_4
\end{array}
\right)
= 
\left(
\begin{array}{c}
0 \\
0 
\end{array}
\right)
\hfill
\end{eqnarray}

which is equivalent to 
\begin{eqnarray}
\left(
\begin{array}{c}
C_{33} n_3 +  C_{34} n_4 \\
C_{43} n_3 +  C_{44} n_4 
\end{array}
\right)
= 
\left(
\begin{array}{c}
-C_{31} n_1  - C_{32} n_2  \\
-C_{41} n_1  - C_{42} n_2   
\end{array}
\right) \,\, .
\end{eqnarray}

~ \\
In matrix form:
\begin{eqnarray}
{\mathbf C_s} \cdot  \vec{n}_s=
\vec{C}_1  \cdot  n_1 +  \vec{C}_2 \cdot n_2  
\end{eqnarray}

where 
\begin{eqnarray}
\vec{n}_s =
\left(
\begin{array}{c}
n_3 \\
n_4 
\end{array}
\right) \, ,
\end{eqnarray}
the reduced matrix for the excited levels is
\begin{eqnarray}
{\mathbf C_s} = 
\left(
\begin{array}{cc}
C_{33} & C_{34} \\
C_{43} & C_{44} 
\end{array}
\right)
\end{eqnarray}
~ \\
and the column vectors  are
\begin{eqnarray}
\vec{C}_1 =
\left(
\begin{array}{c}
-C_{31} \\
-C_{41} 
\end{array}
\right)
\end{eqnarray}
and 
\begin{eqnarray}
\vec{C}_2 =
\left(
\begin{array}{c}
-C_{32} \\
-C_{42} 
\end{array}
\right) \, .
\end{eqnarray}

~ \\
The solution for the excited levels is
\begin{eqnarray}
\vec{n}_s &=&
[{\mathbf C_s}]^{-1} \cdot  \vec{C}_1  \cdot  n_1 + 
[{\mathbf C_s}]^{-1} \cdot  \vec{C}_1  \cdot  n_1 =
\nonumber \\
&=&
\vec{{\cal F}}_1 \cdot  n_1 +  
\vec{{\cal F}}_2 \cdot n_2
\label{eq:ANsol}
\end{eqnarray}
in which the column vectors are
\begin{eqnarray}
\vec{{\cal F}}_1 \equiv  [{\mathbf C_s}]^{-1} \cdot  \vec{C}_1  
\hspace{20pt}\mathrm{and}\hspace{20pt}
\vec{{\cal F}}_2 \equiv  [{\mathbf C_s}]^{-1} \cdot  \vec{C}_2  
\,\, .
\end{eqnarray}

~ \\
The inverse of ${\mathbf C_s}$ is 
\begin{eqnarray}
[{\mathbf C_s}]^{-1} 
=
\frac{1}{C_{33}C_{44} - C_{34}C_{43}} \,
\left(
\begin{array}{cc}
C_{44}  & -C_{34} \\
-C_{43}  &  C_{33} 
\end{array}
\right)
\end{eqnarray}
and 
\begin{eqnarray}
(\mathbf{F}) = 
\left( 
\begin{array}{cc}
\vec{{\cal F}}_1 & \vec{{\cal F}}_2
\end{array}
\right)
\end{eqnarray}
becomes
\begin{eqnarray}
\hspace*{-30pt}(\mathbf{F}) = 
\frac{1}{C_{33}C_{44} - C_{34}C_{43}} \,
\left( 
\begin{array}{cc}
-C_{31}C_{44} + C_{34}C_{41} & -C_{32}C_{44} + C_{34}C_{42} \\
 C_{31}C_{43} - C_{33}C_{41}  &  C_{32}C_{43} - C_{33}C_{42}
\end{array}
\right) \, \, .
\end{eqnarray}
~ \\

The inverse of the reduced matrix is 

\begin{eqnarray}
\hspace*{-70pt}(\mathbf{R}) \equiv (\mathbf{F})^{-1}
=
\frac{1}{C_{31}C_{42} - C_{32}C_{41}} \,
\left( 
\begin{array}{cc}
C_{32}C_{43} - C_{33}C_{42} & C_{32}C_{44} - C_{34}C_{42} \\
-C_{31}C_{43} + C_{33}C_{41}  & -C_{31}C_{44} + C_{34}C_{41}
\end{array}
\right) \, \, .
\end{eqnarray}

~ \\
From the definition (\ref{eq:sxbn}) for the S/XB coefficients
\begin{eqnarray}
{\cal SXB}_{\mu k} \equiv 
\frac{1}{A_{k}}  S_\mu \, {\cal R}_{\mu k } 
\end{eqnarray}
we obtain 

\begin{eqnarray}
\hspace*{-70pt}{\cal SXB}
=
\frac{1}{C_{31}C_{42} - C_{32}C_{41}} \,
\left( 
\begin{array}{cc}
   \frac{S_1}{A_{31}} \left( C_{32}C_{43} - C_{33}C_{42} \right)
 & \frac{S_2}{A_{31}} \left( -C_{31}C_{43} + C_{33}C_{41} \right) \\
   \frac{S_1}{A_{42}} \left( C_{32}C_{44} - C_{34}C_{42} \right)  
 & \frac{S_2}{A_{42}} \left( -C_{31}C_{44} + C_{34}C_{41} \right)
\end{array}
\right)
\end{eqnarray}
~ \\
and replacing the corresponding elements of the matrix (Eq.~(\ref{eq:Ccoeff})), we obtain 
finally the expressions for the S/XB coefficients

\begin{eqnarray}
\hspace*{-70pt}{\cal SXB}_{11} = \frac{n_e S_1}{A_{31}} 
\frac{n_e Q_{23} \, n_e Q_{34}-n_e Q_{24} \, (-A_{31}-A_{32}-n_e Q_{31}-n_e Q_{32}-n_e Q_{34} )}
{n_e Q_{13} n_e Q_{24} - n_e Q_{14} n_e Q_{23} }
 \\
\hspace*{-70pt}{\cal SXB}_{21} = \frac{n_e S_2}{A_{31}} 
\frac{-n_e Q_{13} \, n_e Q_{34} + n_e Q_{14}  (-A_{31}-A_{32}-n_e Q_{31}-n_e Q_{32}-n_e Q_{34} )}
{n_e Q_{13} n_e Q_{24} - n_e Q_{14} n_e Q_{23} }
\nonumber \\
\hspace*{-70pt}{\cal SXB}_{12} = \frac{n_e S_1}{A_{42}} 
\frac{n_e Q_{23} (-A_{41}-A_{42}-A_{43}-n_e Q_{41}-n_e Q_{42}-n_e Q_{43})-n_e Q_{24} (A_{43}+n_e Q_{43})  } 
{n_e Q_{13} n_e Q_{24} - n_e Q_{14} n_e Q_{23} }
\nonumber \\
\hspace*{-70pt}{\cal SXB}_{22} = \frac{n_e S_2}{A_{42}} 
\frac{n_e Q_{13} (A_{41}+A_{42}+A_{43}+n_e Q_{41}+n_e Q_{42}+n_e Q_{43})+n_e Q_{14} (A_{43}+n_e Q_{43})  } 
{n_e Q_{13} n_e Q_{24} - n_e Q_{14} n_e Q_{23} }
\nonumber \, .
\label{eq:SXB4}
\end{eqnarray}

\noappendix
\clearpage
\pagebreak


\end{document}